\documentclass[fleqn,usenatbib]{mnras}

\usepackage{newtxtext,newtxmath}
\usepackage[utf8]{inputenc}
\usepackage{amsmath}

\usepackage[T1]{fontenc}

\DeclareRobustCommand{\VAN}[3]{#2}
\let\VANthebibliography\thebibliography
\def\thebibliography{\DeclareRobustCommand{\VAN}[3]{##3}\VANthebibliography}

\usepackage{graphicx}	
\usepackage{amsmath}	
\usepackage{array, multirow, graphicx, scalefnt}
\usepackage{gensymb}
\usepackage{adjustbox}
\usepackage{multicol}

\makeatletter
\renewcommand*\env@matrix[1][\arraystretch]{%
  \edef\arraystretch{#1}%
  \hskip -\arraycolsep
  \let\@ifnextchar\new@ifnextchar
  \array{*\c@MaxMatrixCols c}}
\makeatother

\title[Evolution of Sub-mm/NIR Sgr A* Images]{The Relationship Between Simulated Sub-Millimeter and Near-Infrared Images of Sagittarius A* from a Magnetically Arrested Black Hole Accretion Flow}

\author[A. Grigorian et al.]{
A. Grigorian,$^{1, 3}$\thanks{E-mail: arpi.grigorian@colorado.edu}
J. Dexter,$^{1,2}$
\\
$^{1}$JILA, University of Colorado and National Institute of Standards and Technology, 440 UCB, Boulder, CO, 80309-0440, USA\\
$^{2}$Department of Astrophysical and Planetary Sciences, University of Colorado, Boulder, 391 UCB, Boulder, CO 80309-0391, USA\\
$^{3}$Department of Physics, University of Colorado, Boulder, 390 UCB, Boulder, CO 80309-0390, USA
}

\date{Accepted XXX. Received YYY; in original form ZZZ}

\pubyear{2023}

\begin{document}
\label{firstpage}
\pagerange{\pageref{firstpage}--\pageref{lastpage}}
\maketitle

\begin{abstract}
Sagittarius A*~(Sgr~A*), the supermassive black hole at the center of the Milky~Way, undergoes large-amplitude near-infrared~(NIR) flares that can coincide with the continuous rotation of the NIR emission region. One promising explanation for this observed NIR behavior is a magnetic flux eruption, which occurs in three-dimensional General~Relativistic Magneto-Hydrodynamic~(3D~GRMHD) simulations of magnetically arrested accretion flows. After running two-temperature 3D~GRMHD simulations, where the electron temperature is evolved self-consistently along with the gas temperature, it is possible to calculate ray-traced images of the synchotron emission from thermal electrons in the accretion flow. Changes in the gas~dominated~($\sigma=b^2/2\rho<1$) regions of the accretion flow during a magnetic flux eruption reproduce the NIR~flaring and NIR~emission region rotation of Sgr~A* with durations consistent with observation. In this paper, we demonstrate that these models also predict that large~(1.5x~-~2x) size increases of the sub-millimeter~(sub-mm) and millimeter~(mm) emission region follow most NIR~flares by~20~-~50~minutes. These size increases occur across a wide parameter space of black hole spin~($a=0.3,0.5,-0.5,0.9375$) and initial tilt angle between the accretion flow and black hole spin axes~$\theta_0$~($\theta_0=0^{\circ}$,~$16^{\circ}$,~$30^{\circ}$). We also calculate the sub-mm polarization angle rotation and the shift of the sub-mm spectral index from zero to~-0.8 during a prominent NIR~flare in our high spin~($a=0.9375$) simulation. We show that, during a magnetic flux eruption, a large~($\sim10r_g$), magnetically dominated~$(\sigma>1)$, low density, and high temperature ``bubble'' forms in the accretion flow. The drop in density inside the bubble and additional electron heating in accretion flow between~15~$r_g$-~25~$r_g$ leads to a sub-mm size increase in corresponding images.

\end{abstract}




\section{Introduction}

Sagittarius A* (Sgr A*), the object at the center of the Milky Way, is a bright radio source~\citep{BalickBrownradiosgrAstar1974, Ekers1975_radiosource, Lo1975_radioSgrAstar} with variable near-infrared (NIR) emission reaching a factor of 10x~-~25x~\citep{genzel_OG_NIR_2003, Ghez2004, GRAVITY2020_NIRFLUXdist} its median value in flux~\citep{Schodel2011, DoddsEden2011, Witzel2018_NIRsurvey}. Sgr A* also exhibits X-ray flares reaching 50x the quiescent value~\citep{Markoff2001} which often coincide with a NIR flare~\citep{Baganoff2001, Genzel_flare_2003, Ghez2004, Eckart2008, Ponti2017, GRAVITY2021_simXray}. Astrometry of stars orbiting Sgr A*~\citep{Ghez2008, Genzel1997} and the resolved black hole shadow at 230~GHz captured by the Event Horizon Collaboration (EHT)~\citep{EHTsgrAstar_1} provide consistent and strong evidence that Sgr A* is a supermassive black hole with mass~$\sim 4 \times 10^6 \odot$. Linear polarization fractions of~$\cong$ 10\%~-~40\% in the NIR and 230~GHz indicate that the emission region around the black hole is dominated by synchrotron radiation~\citep{Eckart2008, Trippe2007, Wielgus2022_polangle}. The lack of sub-mm/mm variability on the same order as the NIR variability suggests that the origin of the flares are electrons accelerated to extreme velocities ($\gamma \gtrsim 10^3$)~\citep{Markoff2001, DoddsEden2009}. For example, the magnetic Rayleigh-Taylor instability stemming from the interface of a dense plasma and a magnetic cavity can drive efficient particle acceleration \citep{Zhdankin2023_RT}. Weaker flares may be also be caused by the lensing of turbulent accretion flow~\citep{Chan2015}. However, the origin of the NIR flares, especially stronger ones, remains unclear.  
 
Recently, the GRAVITY Collaboration measured the movement of SgrA*'s NIR centroid. During three NIR flares~\citep{GRAVITY2018}, the NIR centroid made a clockwise, quasi-circular loop on the sky with a diameter of~$\sim 110 \mu \textrm{as} -140 \mu \textrm{as}$ over periods ranging between 40~minutes~-~70~minutes. Simultaneous measurements of the NIR polarization angle showed continuous rotation -- a clockwise ``loop" in Stokes~$\mathcal{U}, \mathcal{Q}$ space. The lack of a strong Doppler signal from the centroid motion constrained the inclination of the disk to be~$\leq 130^{\circ}$~\citep{GRAVITY2020a}. 

Sgr A* is also known to undergo steady rotations of the sub-mm polarization angle~\citep{Marrone2006_polSgrA}. Using the ALMA Observatory, ~\citet{Wielgus2022_polangle} recently found an example of the 230~GHz polarization angle that makes a quasi-circular path in Stokes~$\mathcal{U}, \mathcal{Q}$ (similar to the path of NIR polarization angle) over a period of 74~$\pm$ 6~minutes following an X-ray flare.

The GRAVITY observations are well described by a hot-spot at~$\simeq 6 r_g - 10 r_g$ orbiting around the black hole at a low ($i \leq 30^{\circ}$) inclination angle and polodial field~\citep{GRAVITY2018,GRAVITY2020}. However, these models do not address possible physical mechanisms that could change the gas and electron temperature in the inner accretion flow.

Three-dimensional General Relativistic Magneto-Hydrodynamic (GRMHD) simulations of a black hole surrounded by an accretion disk or flow offer insights into the physical mechanisms that may be present in Sgr A*. Two key GRMHD black hole accretion disk models often explored in the literature are Standard and Normal Evolution (SANE) disks and Magnetically Arrested Disks (MAD). SANE disks, which do not reach high levels of magnetic flux threading the horizon \citep{Narayan2012_GRMHDSANEsim}, likely rely on the Magnetorotational Instability (MRI) for angular momentum transport ~\citep{HawleyBalbus1991_finite_diff_paper, BalbusHawley1998}. MADs, by contrast, accrete large amounts of magnetic flux that ultimately saturates \citep{BKR1974, BKR1976, Igumenshchev2003_MAD, Begelman2022_MADdisks_def}. 

GRMHD simulations of Magnetically Arrested Disks (MAD) successfully reproduce many observations of Sgr A* \citep[e.g.,][]{EHT5_sgra, Chashkina2021_stripedjets_MADdisks, Dexter2020_survey, Ripperda2020}. MAD simulations of thick disks with high ($a = 0.9375$) black hole spin produce high-temperature orbiting hot-spots roughly consistent with the quasi-circular motion of the NIR centroid observed by GRAVITY \citep{giantGRMHD, Dexter2020_NIRcentroid}.

In this paper, we study similar MAD simulations as \citet{Dexter2020_NIRcentroid} over a wider parameter space. Specifically, we vary the black hole spin $a$ and consider simulations that begin with an initial tilt angle $\theta_0$ between the black hole spin axis and the accretion flow spin axis. In Section~\ref{sec:methods} (Methods), we describe the GRMHD set-up, how images are created using outputted data from our simulations, and how we extract key observables from those images. Then, we focus on an example from the high spin~($a = 0.9375$), prograde ($\theta_0 = 0^{\circ}$) simulation where the dissipation of magnetic flux near the horizon creates a magnetically dominated, low density ``bubble'' (hot spot) within 10~$r_g$ of the accretion flow. The bubble itself is a high $\sigma = b^2/2\rho > 1$ region, where $b^2/2$ is the magnetic pressure and $\rho$ is the fluid density. However, the bubble formation and orbit around the black hole is coincident with significant electron temperature increases in regions where the plasma parameter $\sigma < 1$. When the bubble eventually reaches larger radii (15 $r_g$ - 30 $r_g$), the electron temperature also increases in this region. We show these changes in the $\sigma < 1$ regions of the inner accretion flow not only power a NIR flare and quasi-circular NIR centroid rotation (similar to results from \citet{Dexter2020_NIRcentroid}), but also cause a significant increase in both the sub-mm and mm emission regions about 1~hour after the NIR flare peak. This example produces the rotation of the sub-mm polarization angle with a period of~$\sim$2~hours, consistent with observations \citep{Marrone2006_polSgrA, Wielgus2022_polangle}. The luminosity spectrum evolves during the course of the NIR~flare, and the peak of the sub-mm size increase is simultaneous with a decrease in the synchotron spectrum peak frequency $\nu_c$ and, as a result, a decrease in the spectral index near~230~GHz~(1.3~mm). 

In Section \ref{sec:allsim_fNIR_s230}, we demonstrate that the relationship between NIR flares and the sub-mm/mm~size increases is a feature of all simulations in our parameters space. The relationship between the NIR total flux and the sub-mm/mm size increases is stronger and more consistent that the relationship between the NIR total flux and the sub-mm total flux increases. We also present the NIR centroid behavior during our simulations. Finally, we discuss limitations of the GRMHD model, the dependence of size increases on inclination angle, and the implications our results have on illuminating the physical mechanisms behind NIR flares in Sgr~A*. 

\section{Methods}\label{sec:methods} 


\subsection{The GRMHD MAD Model}\label{sec:GRMHD_info}
 
We carried out 3D GRMHD simulations for a range of black hole spin~$a$ and initial tilt angle of the accretion flow~$\theta_0$ (See Table \ref{tab:sim_names} for the full parameter space). The dimensionless black hole spin is defined as~$a \equiv J/(GM^2/c)$, where~$J$ is the angular momentum and $M$ is the black hole mass. We refer to simulations with $|a| > 0.5$ as high-spin; simulations with $|a| \leq 0.5$ are low-spin. The angle~$\theta_0$ is measured between the spin axis of the black hole and the initial angular momentum vector of the accretion flow. Simulations with~$\theta_0 = 0^{\circ}$ are prograde (the black hole spin and the accretion flow spin are aligned). We use the convention of negative spin values for simulations where the black hole spin and the accretion flow spin are completely anti-aligned. 

Simulations were carried out using the public \texttt{harmpi} code \citep{Tchekhovskoy2019_HARMPI3D}, a 3D version of HARM \citep{Gammie2003_HARM, Noble2006_HARM}. All simulations studied here were initialized from a hydrodynamic equilibrium torus with an inner radius of $12 r_g$ and a pressure maximum at $25 r_g$. The (arbitrary) initial magnetic field was composed of a single poloidal loop so that $\textrm{max}(p_g)/\textrm{max}(p_B) = 100$, where $p_g$, $p_B$ are the gas and magnetic pressures, respectively (See \citet{Tchekhovskoy2011_MAD}). In order to study MAD conditions, the initial magnetic field was concentrated further out in the torus so that magnetic flux could accumulate and saturate on the black hole. For more details, see \citet{Dexter2020_NIRcentroid}.

We use Kerr-Schild coordinates and a grid resolution of 320x256x160 in the~$r$,~$\theta$, and~$\phi$ directions, respectively. The~$r$ values are spaced log-normally to resolve the important physics occurring near the horizon. The $\theta$ coordinate is irregular, chosen to concentrate resolution near the midplane at small radius. The metric in all simulations is fixed as the stationary Kerr metric determined by the black hole mass and spin. We used a temporal cadence of 10 GM/c$^3$ when plotting the fluid variables and observables calculated from ray-traced images as functions of time.

We also include a scheme in our simulations that self-consistently evolves the electron internal energy density for four different electron heating prescriptions along with a single, MHD fluid~\citep{Ressler2015_electronheating}. For this work, we choose the electron energy density calculated from the heating prescription developed by \citet{Werner2018} from their study of particle acceleration from magnetic reconnection using 2D particle-in-cell (PIC) simulations (W18). They found an empirical relationship between the electron heating fraction~$q_e$ and the ion magnetization~$\sigma_i \equiv B^2/4 \pi n_i m_i c^2$, where $B$ is the upstream magnetic field, and $n_i$ is the ion density upstream of the reconnection region~(see their Equation 3). The electron heating fraction ranges from 1/4 for low magnetization to 1/2 for high magnetization. W18 was able to reproduce a variety of observed signatures for Sagittarius A* in previous studies \citep{Dexter2020_NIRcentroid, Dexter2020_survey}. By applying W18 throughout the entire coordinate space of the simulation, our model assumes that 100\% of the dissipated heat is from reconnection.

We evolve our simulations out to times ranging from 20,000~GM/c$^3$ to~100,000~GM/c$^3$~(See Table~\ref{tab:sim_names} for the exact duration for each simulation in our parameter space). The durations are long enough for the fluid in inner ($< 50 r_g$) region to reach equilibrium. For all the $a = 0.9375$ simulations and the $a = 0.5$ simulation with $\theta_0 = 30^\circ$, we ignored the first 10,000~GM/c$^3$ to allow the simulation to evolve from the initial configuration of the fluid to a steady state. However, the remaining low spin $a \leq 0.5$ simulations took longer than 10,000~GM/c$^3$ to start exhibiting consistent magnetic flux eruptions. Our goal is to compare simulations when they are clearly in a MAD state. Therefore, the initial period of time that we ignored for these simulations was longer and ranged from 16,000 GM/c$^3$ to 26,000 GM/c$^3$ (See Table \ref{tab:sim_names}). However, because the simulation durations were long enough, for each simulation there was a significant (> 16,000 GM/c$^3$) continuous period time containing magnetic flux eruptions that we could compare across our parameter space.

\begin{table}
    \caption{A list of all simulations in this paper sorted by spin~$a$ and initial accretion flow tilt angle~$\theta_0$. We ignore times prior to $t_i$ to allow each simulation to evolve from its initial configuration and reach a MAD state. The time $t_f$ is the duration of each simulation. We often refer to the high spin ($a = 0.9375$), prograde simulation as the fiducial simulation in this work.}
    \label{tab:sim_names}
    \begin{tabular}{ lccc }

        \hline
        $a$ & $\theta_0$ & $t_i$~[GM/c$^3$] & $t_f$~[GM/c$^3$] \\
        \hline 
        0.9375 (high-spin) & $0^{\circ}$ (prograde) & 10,000 & 96,310 \\
        0.5 (low-spin) &  $0^{\circ}$ (prograde) & 24,000 & 44,390 \\
        0.9375 (high-spin) & $16^{\circ}$ & 10,000 & 26,660\\
        0.5 (low-spin) & $16^{\circ}$ & 24,000 & 51,710 \\
        0.3 (low-spin) & $16^{\circ}$ & 26,000 & 48,190\\
        0.5 (low-spin) & $30^{\circ}$ & 10,000 & 29,190\\
        -0.5 (low-spin) & $180^{\circ}$ (retrograde) & 16,000 & 34,670\\
    \end{tabular}
\end{table}

\subsection{Synthetic black hole images}\label{sec:create_img_var}

We created monochromatic images for each simulation in Table \ref{tab:sim_names} using the the ray-tracing radiative transfer code \texttt{grtrans}~\citep{Dexter2009, Dexter2016}. The code calculates all four Stokes parameters for each image: total intensity~$\mathcal{I}$, linear polarization~$\mathcal{Q}$ and~$\mathcal{U}$, and circular polarization~$\mathcal{V}$. The \texttt{grtrans} code includes all relativistic effects in a Kerr space time.

We calculate the synchrotron emission with \texttt{grtrans} assuming a thermal distribution of electrons. We do not include emission from non-thermal electrons (see the Discussion for how non-thermal electrons may effect the sub-mm, mm, and NIR emission.) We set the inclination angle~$i$, the angle between the black hole spin axis and the line of sight, to~$i = 25^{\circ}$~(unless otherwise noted). Since the angular momentum of the black hole is along positive~$z$ (spinning counter-clockwise) in our simulation convention,~$i < 90^{\circ}$ means that the black hole, and, generally, the accretion flow, are spinning counter clockwise on the sky. 

When calculating images, we omit regions of the accretion flow which have a $\sigma$ value greater than $\sigma\textrm{-cut}$. Unless specified otherwise, we used $\sigma\textrm{-cut} = 1$. For our fiducial example, we also produced images with $\sigma\textrm{-cut} = 10$ (See Section \ref{eheating_sigma_nonthermal} in the Discussion).

We made images at 3.5~mm~(86~GHz), 1.3~mm~(230~GHz), and 2.2$\mu$m~(136~THz) for all simulations in this work. We refer to all 3.5~mm images as mm (millimeter), 1.3~mm images as sub-mm (sub-millimeter), and 2.2$\mu$m images as NIR (near-infrared). Most images in this study use the following fields of view as a function of frequency: 150$^2$~$\mu \textrm{as}^2$ for the NIR (2.2$\mu$m) images, 300$^2$~$\mu \textrm{as}^2$ for the sub-mm~(1.3~mm) images, and 400$^2$~$\mu \textrm{as}^2$ for the mm~(3.5~mm) images. The black hole is always centered in the field of view. With the exception of images used to calculate the luminosity spectrum, images are 300x300 pixels in size. 

In order to report distances in terms of subtended angles on the sky ($\mu \textrm{as}$) and times in units of~minutes/hours/days, we use 8 kpc for the distance to the Galactic Center and~$M = 4 \times 10^{6} M_{\odot}$ for the mass of the black hole. For reference, 1,000~GM/c$^3$ is approximately 5.6 hrs. 

In GRMHD simulations, the mass accretion rate across the horizon is calculated self-consistently. However, in MAD simulations, matter is accreted in clumps and this ``true '' mass accretion rate is noisy. Instead of using the ``true'' mass accretion rate to scale our images, we choose a constant value for the mass accretion rate \texttt{mdot} every 500 frames so that the average sub-mm total flux is near the observed value of ~$\sim$3~Jy for Sgr A*. When calculating images at higher~$\sigma\textrm{-cut} = 10$, we use a different value of~\texttt{mdot} so that the sub-mm images for each~$\sigma\textrm{-cut}$ meet this observational total flux constraint.

The only exception to the rule above for scaling images is set of $\sigma \textrm{-cut} = 1$ NIR images from the high-spin, prograde simulation between 10,000~GM/c$^3$ and 57,190~GM/c$^3$. For this case, we first found a value of \texttt{mdot} that recreates the observed Sgr A* sub-mm total flux for the first 500 frames and calculated all the NIR images during this period of time with the same choice of \texttt{mdot}. Then, we calculated a ``smoothed'' curve for the mass accretion rate $\langle \dot{M} \rangle$ as a function of time  by taking the actual, intrinsically noisy mass accretion rate across the black hole horizon $\dot{M}$ and smoothing it with a Gaussian kernel. We choose a standard deviation of the Gaussian kernel that produces a $\langle \dot{M} \rangle$ which is monotonically decreasing so that any peaks in $\dot{M}$ do not erroneously create peaks in the NIR total flux curve (we also use this procedure to scale the dimensionless magnetic flux on the black hole horizon, see Section \ref{calc_phiBH}). We then scaled the NIR images by ~$\langle \dot{M} \rangle^2$. The NIR images from the high-spin, prograde simulation after time 57,190~GM/c$^3$ were calculated using the former method. 

\subsection{Calculating Total Flux, Centroid, and Image Size}\label{sec:tf_c_s_calculations}

We calculate the light curves, sizes, and centroid positions directly from the \texttt{grtrans} images. 
 
The moment of an image~$M_{p,q}$ with image coordinates~$x$,~$y$ is defined as
 
 \begin{equation}
 M_{p,q} = \Sigma_x \Sigma_y x^p y^q I(x,y).
\end{equation}

The total flux~$F_{\nu} = \int I_{\nu} \cos(\theta) d\Omega$ from a far-away source can be rewritten as

\begin{equation}
    F_{\nu}  \approx \int I_{\nu}(x, y)\, \textrm{d}x/D \, \textrm{d}y/D
\end{equation}

where~$D$ is the distance to the Galactic Center and~$x$,~$y$ are coordinates on the sky. In this coordinate system, the black hole is centered in the image. Discretizing the integral yields

 \begin{equation}
      F_{\nu} \approx M_{0,0} \, (r_g/D)^2 (X/N) (Y/N)
 \end{equation}
 
 where~$X$,~$Y$ is the total height and width of the field of view in units of the gravitational radius~$r_g$ =~GM/c$^2$ and N is the total number of pixels along the~$x$ and~$y$ directions. 
 
The centroid position on the sky~$C(x,y)$ is given by

\begin{equation}
    C(x,y)=(M_{10}/M_{00}, M_{10}/M_{00}) = (\bar{I_x}, \bar{I_y}).
\end{equation}

To find the size, or the spread of the intensity map, for each image, we first solve for the eigenvalues of the following matrix 

\begin{equation}
\begin{adjustbox}{max width=\columnwidth}$
        \setlength\arraycolsep{1pt}
        \begin{bmatrix}[1.5]
	M_{20}/M_{00} - (M_{10}/M_{00})^2  &  M_{11}/M_{00} - (M_{10}/M_{00})(M_{01}/M_{00}) \\
	M_{11}/M_{00} - (M_{10}/M_{00})(M_{01}/M_{00}) & M_{02}/M_{00} - (M_{01}/M_{00})^2 
	\end{bmatrix}$

\end{adjustbox}
\end{equation}

which gives the major ($\sigma_{MAX}$) and minor ($\sigma_{MIN}$) axes of an ellipse that corresponds to the ``spread'' of the intensity on the image. We approximate the emission region as Gaussian and find the maximum and minimum Full-Width Half-Max ($FWHM_{MAX}$,~$FWHM_{MIN}$, respectively) by using the simple relation

\begin{equation}
    \begin{split}
	FWHM_{MIN} \equiv 2\sqrt{2\ln(2)}\sigma_{MIN} \\
        FWHM_{MAX} \equiv 2\sqrt{2\ln(2)}\sigma_{MAX}.
     \end{split}
\end{equation}

\subsection{The Cross Correlation Coefficient}\label{sec:calc_cc}

 Each of the observables defined above at a single frequency, such as the total flux,~$\bar{I_x}$,~$\bar{I_y}$,~$FWHM_{MAX}$,~$FWHM_{MIN}$, produces a curve that is a discretized function of time with equal time steps: a time series.  The \textit{cross-correlation} of two time series is a convolution of one curve with the other that measures their degree of similarity as well as any characteristic lags between them. 

We use the following definition of a discrete cross-correlation coefficient~$\mathcal{CC}$

\begin{equation}
    \begin{aligned}[b]
        & \mathcal{CC}[A(t), B(t)](\tau) = \\
        & \cfrac{\Sigma_{i = 0}^{N - \tau} (A(i) - \langle B_{0,N - \tau}\rangle) (A(i + \tau) - \langle B_{\tau, N}\rangle )}{\sqrt{\Sigma_i^{N} (A(i) - \langle A_{0, N}\rangle)^2}\sqrt{\Sigma_i^{N}(B(i) - \langle B_{0, N})\rangle )^2}}.
    \end{aligned}
\end{equation}

The cross correlation~$\mathcal{CC}[A(t), B(t)](\tau)$ is the functional that takes two curves~$A(t), B(t)$ and outputs a function that depends on the separation time delay~$\tau$ between the two curves. The quantities~$\langle A_{j,k} \rangle$ and~$\langle B_{j,k} \rangle$ each correspond to the mean value to their respective functions~$A$,$B$ in the interval ($j$,~$k$). The maximum of~$\mathcal{CC}(\tau)$ shows the strength of the correlation between two curves and the location of the peak value is a characteristic delay time~$\tau^{*}$.

Cross correlation values greater than 0.5 are strong, between 0.25 and 0.5 are weak, and values less than 0.25 are practically uncorrelated.  

We can use the cross correlation function to define the auto correlation for a single time series~$A(t)$: ~$\mathcal{CC}[A(t), A(t)](\tau)$. The auto-correlation function measures the periodicity of a time series; a strong auto-correlation coefficient at a non zero ~$\tau^{*}$ suggests the curve contains a mode that has frequency ~$1/\tau^{*}$.

\subsection{The Structure Function}

The first order structure function for a finite, continuous time series~$\mathcal{A}(t)$ is defined as \citep{Simonetti1985, Hughes1992}

\begin{equation}\label{eq:SF_anal}
    \mathcal{SF}(\tau) = \sqrt{\frac{1}{T} \int \, dt \, (\mathcal{A}(t) - \mathcal{A}(t + \tau))^2 }
\end{equation}

where~$T$ is the total duration for the time series. We then find the discrete structure function

\begin{equation}\label{eq:SF}
    SF(\tau_j) = \sqrt{ \sum_{i=1}^{N - j} (F(t_i) - F(t_i + \tau_j))^2/N},
\end{equation}

where~$N$ is the total number of points in our discrete time series~$F(t_i)$, and apply it to our discrete time series that represent simulated observables. 

The structure function ignores any DC offsets present in a time series \citep{Simonetti1985, Dexter2014, Witzel2018_NIRsurvey}. Therefore, we can compare the simulated structure function to observation even if there is a source of steady emission in Sagittarius A* or background emission that is not included in our model. For curves that are well modeled by a Damped Random Walk (DRW), the structure function reaches an asymptotic value~$SF_{\infty}$ at late times \citep{MacLeod2010}. If the structure function becomes close to~$SF_{\infty}$ at finite time ~$\tau_{SF}$, this time represents when the series becomes uncorrelated or is transitioning from red noise to white noise. We choose ~$\tau_{SF}$ as the time when the structure function $SF$ reaches its first prominent peak before plateauing. 

\subsection{Computing Average Values of Fluid Variables}

In this work, most of our plots of fluid variables show averages of that variable over the polar angle $\theta$ or the azimuthal angle $\phi$. For some function $f(r, \theta, \phi)$ that represents a scalar field of the fluid, the polar and azimuthally averaged values $\langle f(r, \phi) \rangle$ and $\langle f(r, \theta) \rangle$ are, respectively

\begin{equation}\label{averages}
    \langle f(r, \phi) \rangle = \frac{\int_{\theta = 0}^{\theta = \pi} \textrm{d}\theta \, \sqrt{-g} f (r, \theta, \phi)}{\int_{\theta = 0}^{\theta = \pi} \textrm{d}\theta \, \sqrt{-g}}
\end{equation}
\begin{equation}    
    \langle f(r, \theta) \rangle = \frac{\int_{\phi = 0}^{\phi = 2\pi} \textrm{d}\phi \, \sqrt{-g} f (r, \theta, \phi)}{\int_{\phi = 0}^{\phi = 2\pi} \textrm{d}\phi \, \sqrt{-g}}
\end{equation}

Sometimes it is more instructive to show the weighted average of a fluid variable with respect to the density $\rho$. The weighted averages over $\theta$ and $\phi$ are given, respectively, as

\begin{equation}
    \langle \tilde{f}(r, \phi) \rangle = \frac{\int_{\theta = 0}^{\theta = \pi} \textrm{d}\theta \, \sqrt{-g} \rho (r, \theta, \phi) f (r, \theta, \phi)}{\int_{\theta = 0}^{\theta = \pi} \textrm{d}\theta \, \sqrt{-g}\rho (r, \theta, \phi)}
\end{equation}
\begin{equation} \label{weighted_averages}  
    \langle \tilde{f}(r, \theta) \rangle = \frac{\int_{\phi = 0}^{\phi = 2\pi} \textrm{d}\phi \, \sqrt{-g} \rho (r, \theta, \phi) f (r, \theta, \phi)}{\int_{\phi = 0}^{\phi = 2\pi} \textrm{d}\phi \, \sqrt{-g} \rho (r, \theta, \phi)}
\end{equation}

\subsection{Calculating the Dimensionless Magnetic Flux on the Event Horizon}\label{calc_phiBH}

As the simulation evolves, magnetized material accretes onto the black hole. This process allows the magnetic flux threading the surface just outside the event horizon to accumulate. The total magnetic flux ~$\Phi$ at over the spherical shell at radius~$R$ is defined by

\begin{equation}
    \Phi = \sqrt{4\pi} \int_{r = R} d\theta d\phi \, B^r \sqrt{-g}
\end{equation}


where~$B^r$ is the radial component of the three-vector magnetic field and~$\sqrt{-g}$ is the determinant of the metric. 

Instead of $\Phi$, we use the \textit{dimensionless} magnetic flux $\phi_{BH}$ given by  

\begin{equation}\label{eq:mag_flux}
\phi_{BH} = \Phi_{BH}/\sqrt{\langle \dot{M} \rangle}.
\end{equation}

We calculate smoothed mass accretion rate $\langle \dot{M} \rangle$ by taking the actual, intrinsically noisy mass accretion rate across the black hole horizon $\dot{M}$ and smoothing it with a Gaussian kernel. We set the standard deviation of the Gaussian kernel so $\langle \dot{M} \rangle$ is monotonically decreasing; this step ensures that the variability of $\dot{M}$ does not translate to the dimensionless magnetic flux curve (this is similar to the procedure used to scale the total flux of some of our NIR images). Therefore, all the fluctuations seen in $\phi_{BH}$ are due to changes in the magnetic flux rather than due to variability in the mass accretion rate.
 
\section{Magnetic Flux Eruptions in the Magnetically Arrested Accretion Flow}
\label{sec:flux_erp_GRMHD} 

For all simulations studied, the magnetic flux on the horizon $\phi_{BH}$ enter periods of slow build up followed by rapid dissipation. Most drops in $\phi_{BH}$, called~\textit{magnetic flux eruptions}, are also accompanied by significant increases in the NIR total flux: a NIR flare.

\begin{figure*}
    \includegraphics[width = \textwidth]{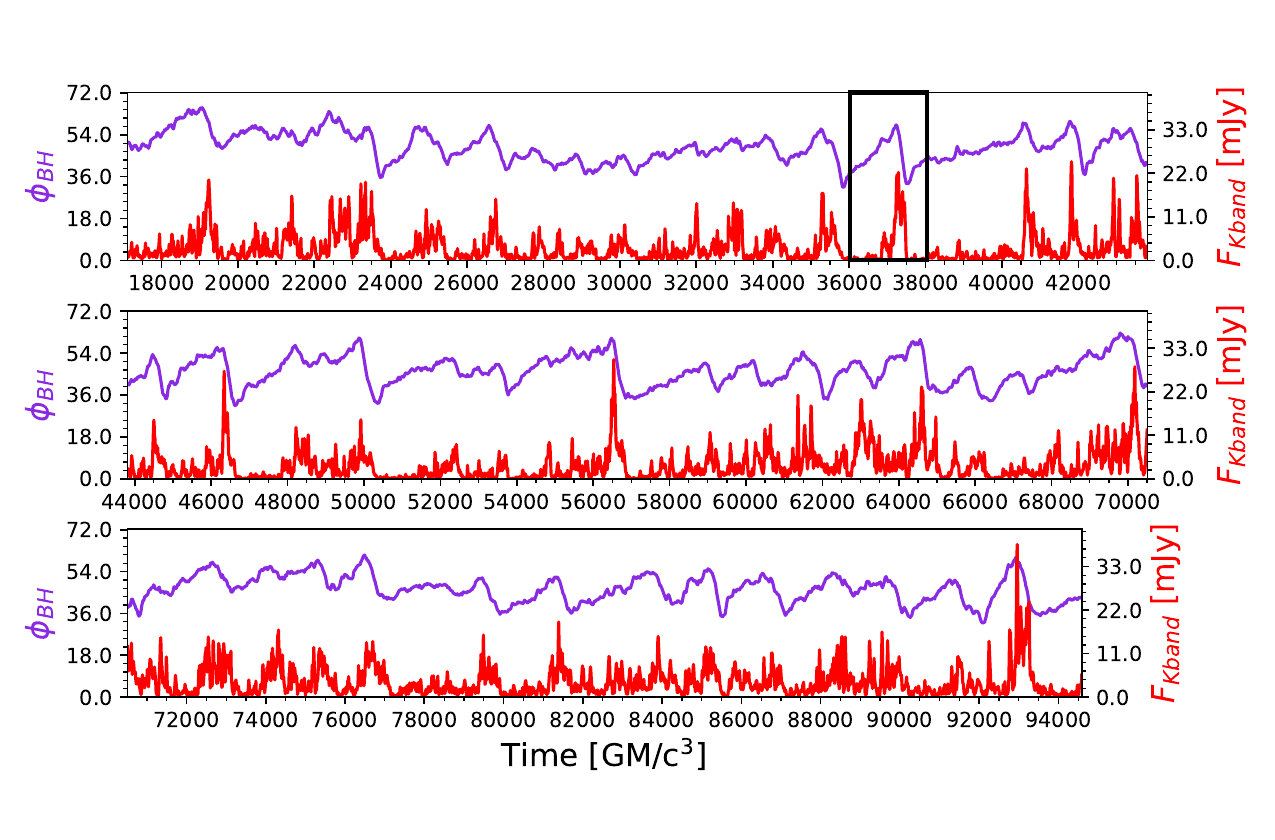}
    \caption{Dimensionless magnetic flux on the horizon~$\phi_{BH}$ (purple) and the NIR total flux curve (red) as a function of time for the high-spin ($a = 0.9375$) prograde simulation. The magnetic flux goes through many cycles of slow buildup followed by rapid dissipation. The dissipation of~$\phi_{BH}$, called a magnetic flux eruption event, coincides with a sharp peaks in the NIR total flux curve.}
    \label{fig:phiBH_NIRtflux}
\end{figure*}

Figure~\ref{fig:phiBH_NIRtflux} shows that $\phi_{BH}$ cycles many times during the high-spin, prograde simulation. Most significant drops in magnetic flux are also accompanied by NIR flares. We find the average time between magnetic flux eruptions to be roughly $\sim$1,600~GM/c$^3$ ($\sim$9~hours), although the low (0.2) auto-correlation peak of the $\phi_{BH}$ curve suggests no consistent period of recurrence for the magnetic flux eruptions. 

\begin{figure*}
    \includegraphics[width = \textwidth]{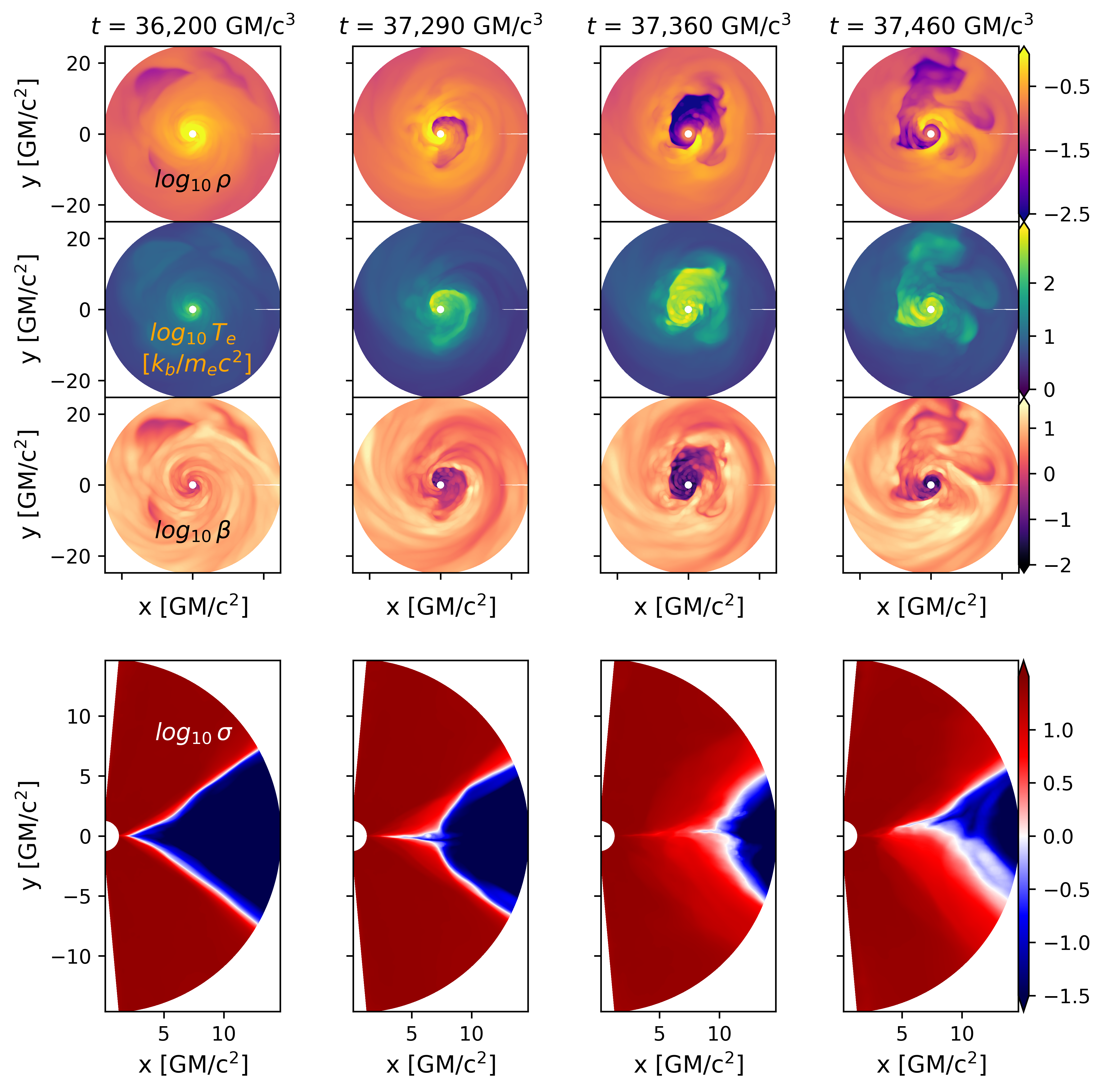}
    \caption{The density~$\rho$ (in code units), the temperature in units of~$kT/m_e c^2$ weighted by the density~$\rho$, and the plasma~$\beta$ parameter~$2p/b^2$, where~$p$ is the fluid pressure and $b^2/2$ is the magnetic pressure, all averaged over polar angle $\theta$. We also show the plasma $\sigma$ parameter $b^2/2\rho$ averaged over azimuthal angle $\phi$ (see Equations \ref{averages} - \ref{weighted_averages} for how we compute averages). Note that the outermost radius shown in the $\sigma$ parameter plots is less than the outermost radius shown for the other fluid parameter plots. A low-density, high temperature bubble forms in the innermost accretion flow and breaks apart before moving to larger radii during the 2,000 GM/c$^3$ interval outlined by a black box in Figure \ref{fig:phiBH_NIRtflux}. In the far left column, the accretion flow is in a quiescent state. The NIR total flux is near its median value and the sub-mm size is~$\sim$50~$\mu \textrm{as}$. The temperature and density is high close to the event horizon at $\sim 2 r_g$ and then falls off at larger radii. Spiral structure in the density plot is visible. At~$t$~=~37,290~GM/c$^3$, a large scale (10~$r_g$) low density, high temperature, and low plasma~$\beta$ region forms.  Also, at this time, the magnetic flux is expelled from the horizon and NIR total flux curve is at a maximum. At~$t$~=~37,360~GM/c$^3$ the low density bubble has increased in size and pushed material out to larger radii. The temperature in this region stays elevated and additional smaller, even higher temperature spots appear throughout the bubble. In the rightmost column, at~$t$~=~37,450~GM/c$^3$, the sub-mm emission region size is at a maximum. The bubble is dissipating. The temperature in the bubble has started to cool and the density of the bubble has increased. There is a small (10x) increase in temperature in the 15~$r_g$ - 25~$r_g$ region. Following the trajectory of the bubble in the last three columns, one can see that it rotates counter-clockwise with the general accretion flow.}
    \label{fig:HARM_var}

\end{figure*}

Figure~\ref{fig:HARM_var} shows the evolution of several fluid variables during the course of a magnetic flux eruption taken from the high-spin, prograde simulation (all times occur within the time interval outlined by a black box in Figure \ref{fig:phiBH_NIRtflux}). 

Shortly after the magnetic flux peak at 37,230~GM/c$^3$, a low density, high temperature ``bubble'' begins to form in the accretion flow. At 37,290~GM/$c^3$, the NIR total flux curve reaches the most prominent peak during this magnetic flux eruption event. The bubble is now $\sim 10 r_g$ in size and the electron temperature in the bubble is 100x - 1000x the median value (second column in Figure~\ref{fig:HARM_var}). The fluid parameter $\sigma = b^2/2\rho$ has also increased above and below the equatorial plane; only a thin region within $\sim 7 r_g$ of the accretion flow remains mostly $\sigma < 1$.

As the bubble rotates around the black hole, it continues to grow. By 37,360 GM/c$^3$, the bubble is $\sim 15 r_g$ in size (third column in Figure~\ref{fig:HARM_var}). At this time, the innermost ($\lesssim 10 r_g$) region of the accretion flow is mostly magnetically dominated. The sub-mm size is also increasing (See Figures \ref{fig:a9375_prograde_example} and \ref{fig:images}).


Around 37,400 GM/c$^3$, the low density region becomes elongated; it stretches to radii $\sim 20 r_g$. Density plots of the slice along the equatorial plane show that a single, low density region breaks apart into multiple low density regions. By 37,460 GM/c$^3$, the slice of the equatorial plane also shows that the low density region within $\sim 10 r_g$ remains magnetically dominated but the $\sigma$ value of the low density region at larger radii has decreased to less than one. Also, at 37,460 GM/c$^3$, the electron temperature has increased to 10x the median value in the 15$r_g$ - 25$r_g$ region of the accretion flow. The temperature within $\sim 10 r_g$ remains elevated (last column of Figure \ref{fig:HARM_var}). The sub-mm size at 37,460 GM/c$^3$ is at a maximum (See Figure \ref{fig:images} for the sub-mm image corresponding to this time).

Throughout all the events described above, the magnetic flux has been monotonically decreasing. The magnetic flux reaches a minimum at 37,540~GM/c$^3$. By this time, the outer region of the bubble has reached 30 $r_g$. The sub-mm size is still elevated from its median value but is decreasing.

The total time between the magnetic flux peak and its local minimum is 310~GM/c$^3$, or 1.7~hours.

\section{NIR Total Flux and Sub-mm/mm Total Flux and Size During a Magnetic Flux Eruption}\label{sec:ex_a9375_cNIR_s86_s230}

\begin{figure*}
    \centering
    \includegraphics[width = \textwidth]{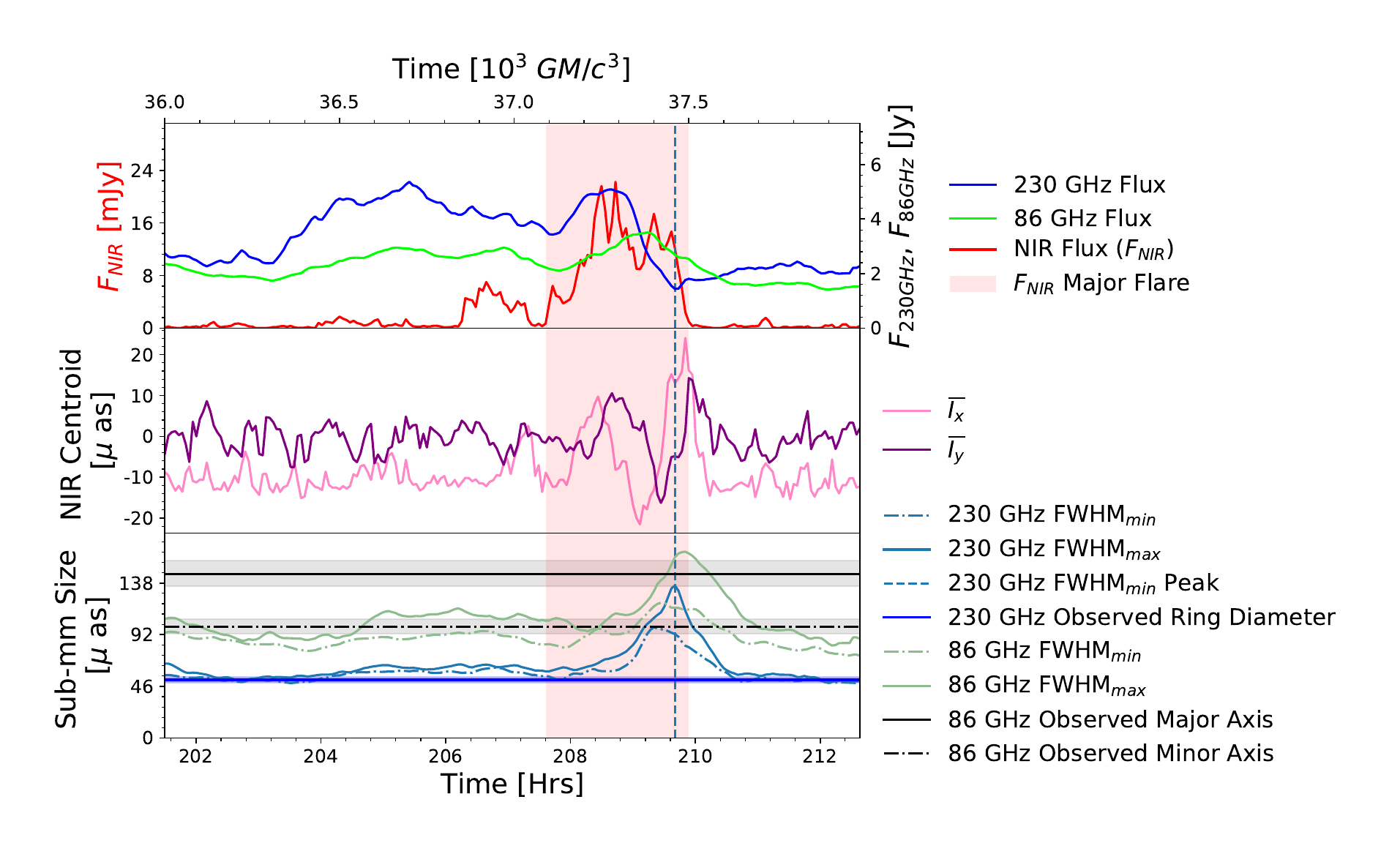}
    \caption{The evolution of key NIR, sub-mm, and mm observables before and during a prominent NIR flare. The 2,000 GM/c$^3$ interval of time shown here corresponds to the black box outlined in Figure \ref{fig:phiBH_NIRtflux} and is taken from the high-spin, prograde simulation. In this example, the significant NIR flare results in both quasi-circular motion of the NIR centroid and an increase in the~$FWHM_{MAX}$ and $FWHM_{min}$ of the sub-mm/mm emission region~(See Section~\ref{sec:tf_c_s_calculations} for the definition of~$FWHM_{MAX}$,~$FWHM_{MIN}$ and the centroid position~$(\bar{I}_x, \bar{I}_y)$). The light red shaded regions corresponds to the second, prominent NIR flare; the dotted, vertical blue line represents the sub-mm~$FWHM_{MAX}$ peak time. Following a period of quiescence, the NIR light curve first exhibits a small,~$\sim$6~mJy flare near 36,900~GM/c$^3$ before a second, larger (22.2~mJy) flare that reaches its peak value at~$t$~=~37,290~GM/c$^3$. The first flare corresponds to a minor eruption of magnetic flux whereas the larger flare occurs during a more dramatic magnetic flux eruption event. Prior to the second flare, both the NIR centroid (middle subplot) and the sub-mm/mm size (lower subplot) undergo only small deviations from their median values. However, during the second flare, the NIR centroid changes its behavior and makes a quasi-circular orbit around the black hole. The sub-mm and mm sizes begin to climb. The sub-mm peak size increase at 37,460~GM/c$^3$ coincides with a final, lesser NIR total flux peak during the second flare before the total flux quickly drops to its median value. The mm~$FWHM_{MAX}$ reaches at maximum only 10~minutes after the sub-mm~$FWHM_{MAX}$ peak and outside the flaring period. Then, the sub-mm and mm sizes slowly decrease back to their quiescent values. The entire event, from the beginning of the first flare to the end of the sub-mm/mm size increases, lasts 800~GM/c$^3$, or just under 4.5~hours. However, the separation between the peak value of the prominent NIR flare and the sub-mm~$FWHM_{MAX}$ peak is only 57~minutes (270~GM/c$^3$). The width of the sub-mm size increase peak is less than~2~hours. We also include the observed sub-mm size from the EHT~\citep{EHTsgrAstar_1} and the observed mm sizes from ALMA~\citep{Issaoun2021} in the bottom panel of the plot. The light blue region around the blue EHT line represents the 68\% credible interval; the light grey region around the observed mm~$FWHM_{MAX}$/$FWHM_{min}$ lines represents the 95\% confidence interval for each measurement. We discuss the agreement between the sub-mm size and our model, as well as the tension between our calculated mm sizes and the observed values, in Section~\ref{discuss:sizes} of the Discussion.}
    \label{fig:a9375_prograde_example}
\end{figure*}

\begin{figure*}
	\includegraphics[width  = .49\textwidth]{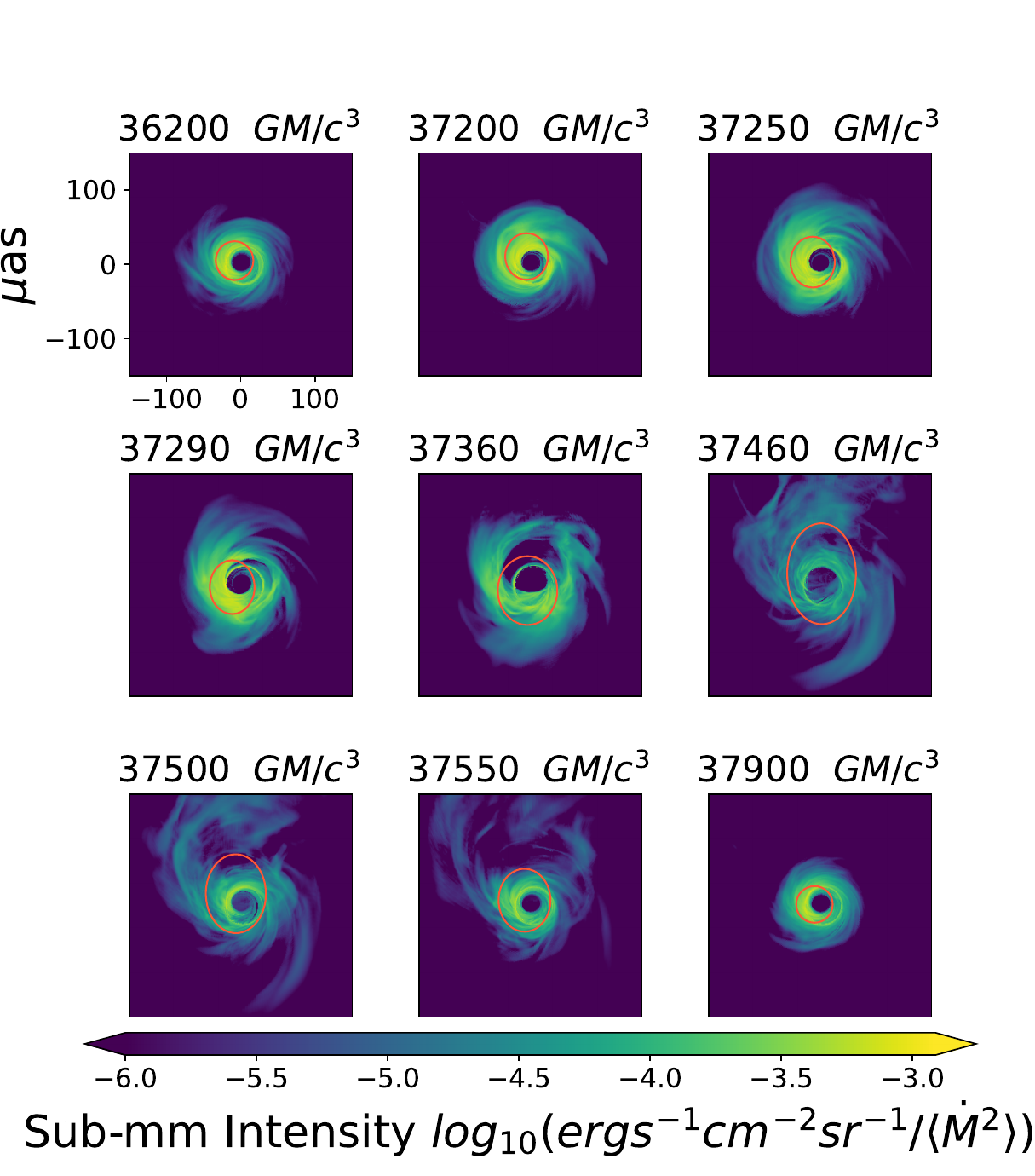}
	\includegraphics[width = .49\textwidth]{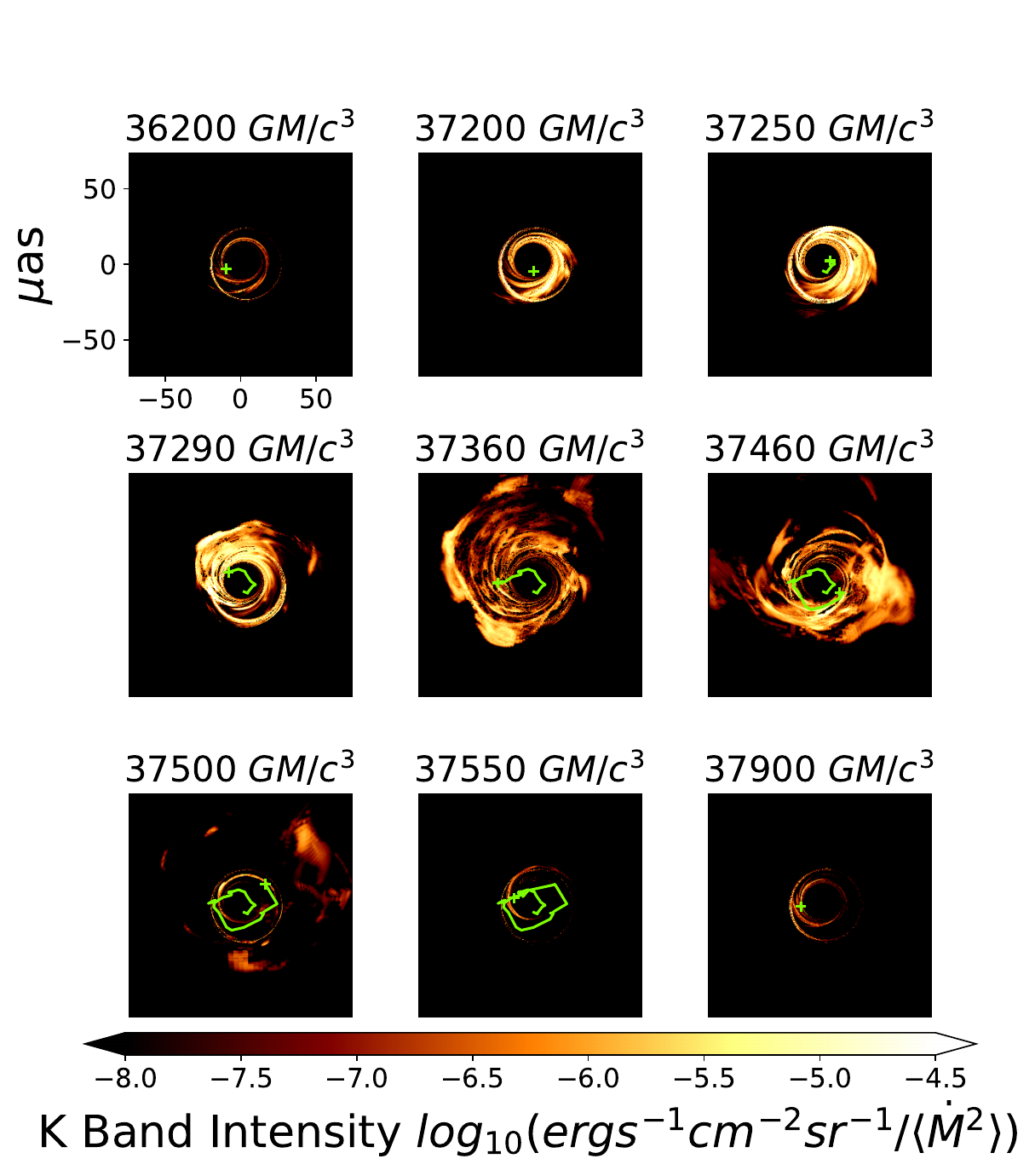}
	\caption{Select sub-mm~(left) and NIR~(right) images from the high-spin, prograde simulation that includes the example NIR flare. All times occur during the 2,000 GM/c$^3$ interval shown in Figure \ref{fig:a9375_prograde_example}. The NIR images are squares of 30~$GM/c^2$ (150~$\mu\textrm{as}$ on the sky), the sub-mm images are squares of size 60~$GM/c^2$ (300~$\mu\textrm{as}$ on the sky), and the black hole is centered at the origin in both sets of images. The minor and major axes of the red ellipse in the sub-mm correspond sub-mm~$FWHM_{MIN}$ and~$FWHM_{MAX}$, respectively. The green + in the NIR images is the NIR centroid position at the time of the image and past locations of the centroid are represented with a green line. First, a gap in the sub-mm intensity map forms that corresponds to the region of increased NIR emission. The NIR centroid also begins its first, counter-clockwise orbit (from $t$~=~37,200~GM/c$^3$ to $t$~=~37,290~GM/c$^3$) around the black hole. At time 37,360~GM/c$^3$, the gap in the sub-mm intensity map is extremely pronounced and the sub-mm size continues to increase. The NIR intensity in the region 20 $\mu \textrm{as}$ - 50 $\mu \textrm{as}$ from the black hole center also increases, pushing the NIR centroid further away from its median value and starting a second, larger orbit around the black hole. By 37,460~GM/c$^3$, when the sub-mm emission size is at a maximum, the gap in the sub-mm emission has dissipated but the emission in the 50 $\mu \textrm{as}$ - 100 $\mu \textrm{as}$ has increased whereas the emission closer to the black hole center has decreased. As the sub-mm size decreases to its median value during times 37,500~GM/c$^3$ to 37550~GM/c$^3$, the sub-mm intensity near the center of the black hole increases and the extra emission in the outer regions dissipates. The NIR total flux drops significant during this time, but the NIR centroid makes it back to its median value by 37550~GM/c$^3$. By 37,900~GM/c$^3$, both the sub-mm size and NIR centroid have returned to their median values.}
    \label{fig:images}
 \end{figure*}

During the course of our fiducial magnetic flux eruption explored above (Section \ref{sec:flux_erp_GRMHD}), the NIR flux significantly increases or ``flares''. It reaches a peak value of 22.2~mJy, or 10x of the median NIR flux, at 37,290 GM/c$^3$ (Figure \ref{fig:a9375_prograde_example}). After the NIR flux peak, the NIR total flux stays elevated and reaches several lesser peaks over the course of roughly 1 hour before quickly returning to its median value.

This NIR flare is also accompanied by significant ($\sim$2x) increases in both the sub-mm and mm $FWHM_{MIN}$ and~$FWHM_{MAX}$. Both the sub-mm and mm size reach a peak value about 1~hour after the NIR total flux peak. The mm~$FWHM_{MAX}$ peak lags behind the sub-mm~$FWHM_{MAX}$ peak by 10~minutes. Near the peak value of the sub-mm and mm size increases, the NIR total flux curve quickly drops back to its median value. The sub-mm/mm sizes, however, slowly return to their quiescent values over the course of ~$\sim$30~minutes.

Before the NIR flare, the sub-mm total flux shows significant variability between 2~Jy to 5~Jy. During the NIR flare, the sub-mm flux climbs again to 5~Jy. Therefore, the peak in sub-mm flux is not unique to periods with NIR flaring. 

After the second 5~Jy peak, the sub-mm flux quickly drops below 2~Jy. The minimum in the sub-mm total flux curve is simultaneous with the sub-mm~$FWHM_{MAX}$ peak as well as a minor peak in the NIR total flux curve~(See dotted vertical blue line in Figure~\ref{fig:a9375_prograde_example}).

The mm total flux, both during the quiescent and flaring periods, shows less variability than its sub-mm counterpart and stays near 2.5~Jy. 

Figure~\ref{fig:a9375_prograde_example} also shows a minor NIR flare near 36,800~GM/c$^3$, which is coincident with a minor magnetic flux eruption event. The NIR flux during the flare only reaches a peak value~$\sim$6~mJy and stays elevated for~80~minutes. Also, unlike the prominent NIR flare, there are no significant sub-mm/mm size increases within 1 hour after the minor NIR flare. The low density ``bubble'' that forms during the minor flare is smaller in size than its counterpart from the prominent flare, demonstrating that the geometry of the bubble has an impact on the resulting sub-mm/mm size. 

\section{The Sub-mm and NIR Centroid Motion During a Magnetic Flux Eruption}

\subsection{The NIR Centroid}

Prior to the prominent magnetic flux eruption discussed above from the high-spin, prograde simulation, the NIR centroid curves stay near the median value of (-6.5~$\mu \textrm{as}$, 0.3~$\mu \textrm{as}$). The NIR centroid position does not change significantly during the minor NIR flare (Figure \ref{fig:a9375_prograde_example}).

However, during the second, prominent NIR flare, the NIR centroid makes two orbits around the black hole over a period of $\sim$ 3~hours, tracing a quasi-circular path with a peak diameter of 46~$\mu$as (See the right side of Figure~\ref{fig:images} and Figure \ref{fig:a9375_prograde_example}). The NIR centroid returns to its quiescent position at the end of the NIR flare.

\subsection{Sub-mm Centroid Motion and Linear Polarization}\label{sec:lin_pol}

\begin{figure*}
    \centering
    \includegraphics[width = \textwidth]{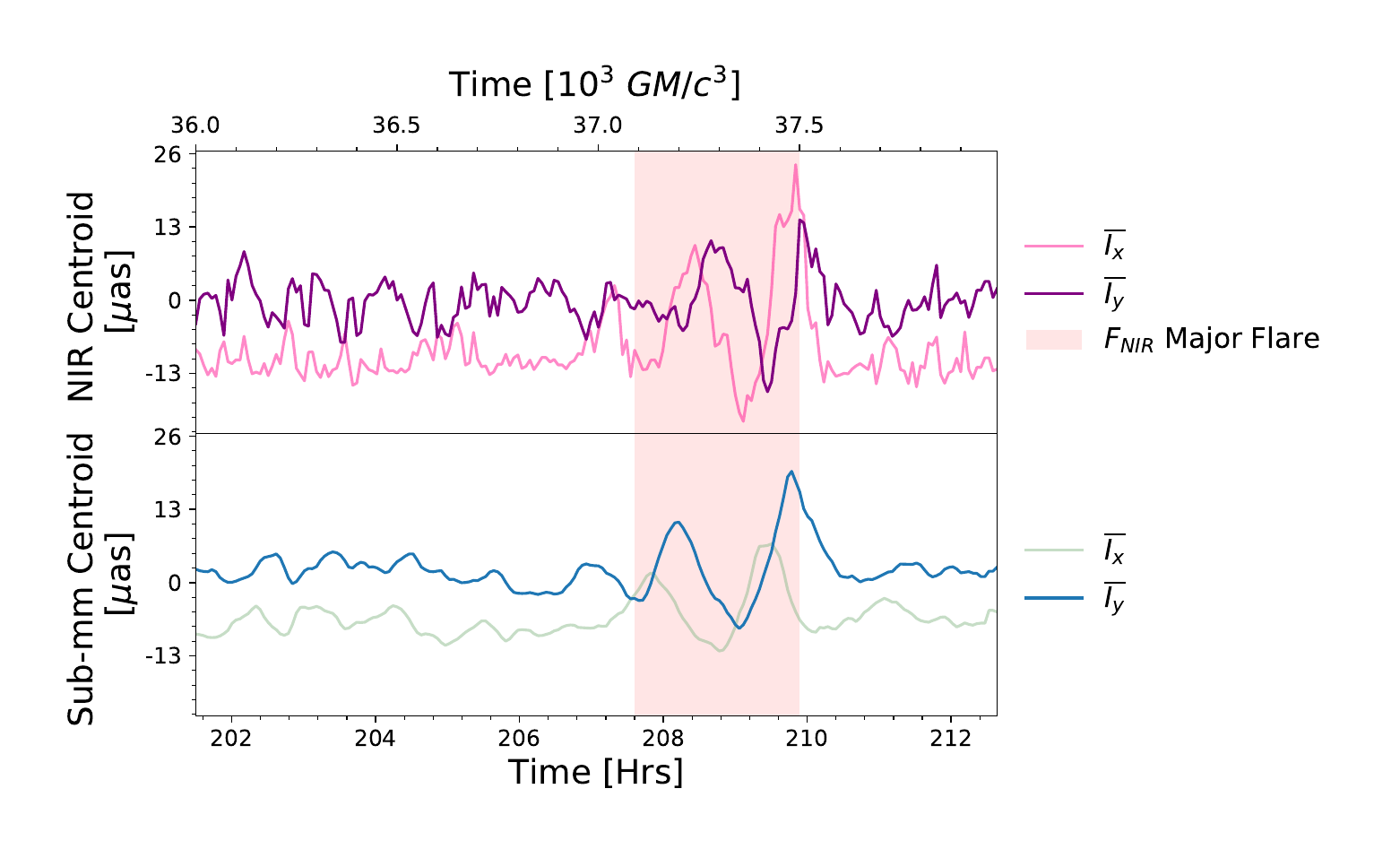}
    \caption{NIR and sub-mm centroid motion from the 2,000 GM/c$^3$ example which includes the prominent NIR flare, quiescent periods, and a minor flare. Both centroids stay close to their median values during quiescent periods and the first (minor) flare. However, during the second, more prominent flare, both the sub-mm and NIR centroids exhibit similar quasi-circular motion. The NIR centroid lags behind its sub-mm counterpart by $\sim$~20~minutes.} 
    \label{fig:a9375_prograde_example_centroids}
\end{figure*}

During the prominent NIR example flare, the sub-mm centroid makes a similar, quasi-circular orbit on the sky as the NIR centroid over a period of $\sim$ 3~hours (Figure \ref{fig:a9375_prograde_example_centroids}). The sub-mm centroid position is also less noisy than its NIR counterpart. Considering both centroids follow the counter-clockwise motion of the accretion flow, the sub-mm centroid is ahead of the NIR centroid by $\sim 20$ minutes. The effective diameter of the sub-mm centroid path, which reaches an extreme at 28~$\mu$as, is also smaller than its NIR counterpart.

\begin{figure}
    \centering
    \includegraphics[width=\columnwidth]{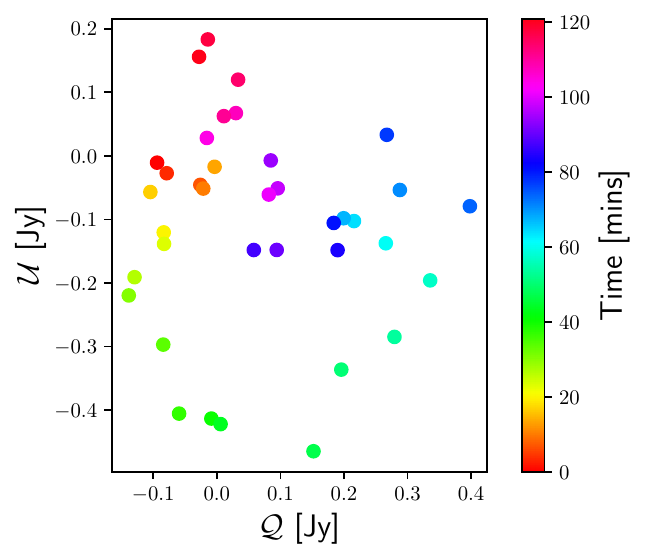}
    \caption{Integrated Stokes~$\mathcal{Q}$,~$\mathcal{U}$ for each time step during the example NIR flare explored in Section \ref{sec:ex_a9375_cNIR_s86_s230} and shown in Figures \ref{fig:a9375_prograde_example}, \ref{fig:images}, and \ref{fig:a9375_prograde_example_centroids}. The linear polarization slowly rotates in Stokes~$\mathcal{U}$ and~$\mathcal{Q}$ space over a period of~2~hours in a quasi-circular pattern. In this example, polarization angle rotation is consistent with quasi-circular centroid motion at the same frequency on the sky.} 
    \label{fig:example_totQ_totU}
\end{figure}

During the 5~hours prior to the flare, we find that the sub-mm linear polarization integrated over the image behaves like a random walk in Stokes~$\mathcal{U}$ and~$\mathcal{Q}$ space with an average linear polarization fraction ($p = (|\mathcal{U}|^2 + |\mathcal{Q}|^2)/|\mathcal{I}|^2$) of 4.5\%. However, right as the sub-mm centroid begins its quasi-circular path, the sub-mm polarization vector changes behavior and slowly rotates in Stokes~$\mathcal{U}$ and~$\mathcal{Q}$. For the first~$\sim$2~hours, shown in Figure \ref{fig:example_totQ_totU}, the path traced in Stokes~$\mathcal{U}$ and~$\mathcal{Q}$ space is a clear loop; the shape of the path in Stokes~$\mathcal{U}$ and~$\mathcal{Q}$ looses this character for the final $\sim$~1 hour of the sub-mm centroid motion. While the polarization vector angle slowly rotates, the total linear polarization fraction reaches a maximum value of 11\%.

\section{The Luminosity Spectrum and Sub-mm Spectral Index During a Magnetic Flux Eruption}\label{sec:spectrum}

We calculate the spectrum~$\nu L(\nu)$ directly from the intensity maps by finding the total flux~$F_{\nu}$ over a wide range of frequencies $\nu$. Then, we find ~$L_{\nu} = 4\pi D^2 F_{\nu}$, where~$D$ is the distance from the galactic center to the observer.

The spectral index is defined as $\alpha \equiv - d\log(F(\nu))/d\log \nu$. We calculate the spectral index near 230~GHz (1.3~mm),~$\alpha_{submm}$, by finding $\log(F(\log(\nu)))$ for five frequencies between 210~GHz and 250~GHz and then fitting the function to a line using \texttt{numpy.polyfit}.

\begin{figure}
    \centering
    \includegraphics[width=\columnwidth]{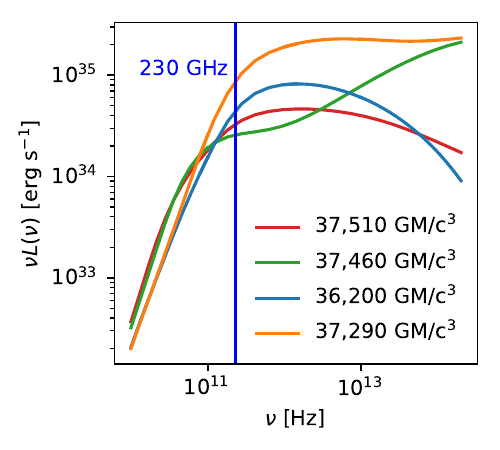}
    \caption{Luminosity spectra for the high-spin, prograde simulation at key times before, during, and after the example NIR flare. At~$t$~=~36,200~GM/c$^3$, the accretion flow emission is in quiescence. The spectrum is flat at low frequencies and falls exponentially at high frequencies. During the peak value of the NIR total flux flare ($t$~=~37290~GM/c$^3$), the spectrum develops an elevated high frequency tail but is only slightly elevated at sub-mm and mm frequencies; ~$\alpha_{submm}$ remains near zero. However, during the peak sub-mm size, $\alpha_{submm}$~=~-0.84. A time steps later, at 37,510~GM/c$^3$, the NIR flux is no longer elevated but the size is sill 1.6x the median value and $\alpha_{submm}$ remains negative. The sub-mm size increase is coincident with a shift in the synchrotron spectrum to lower frequencies and a more negative spectral index near 230~GHz.}
    \label{fig:example_spectrum}
\end{figure}

Figure~\ref{fig:example_spectrum} shows the resulting spectra. The spectrum at~$t$~=~36,200~GM/c$^3$, when the accretion flow is an a quiescent state, has a shape similar to the power spectrum of a single electron and ~$\alpha_{submm}$~=~-0.07. At the NIR flare peak ($t$~=~37,290~GM/c$^3$), the high frequency tail of~$\nu L(\nu)$ is significantly elevated with only a slight increase in luminosity for sub-millimeter and millimeter frequencies but $\alpha_{submm}$ remains relatively unchanged with a value of~0.06. There is also little to no change in the total luminosity for millimeter and radio frequencies.

When the simulated images of the accretion flow undergo sub-mm and mm size increases, the luminosity spectrum continues to evolve. At ($t$~=~37,460~GM/c$^3$), when the sub-mm size reaches its peak value, the spectrum becomes a combination of an elevated high frequency tail and a synchrotron spectrum with a lower peak frequency $\nu_c$ than at the earlier, quiescent time. The shift in the peak synchrotron frequency is clearer at time~$t$~=~37510~GM/c$^3$, when the sub-mm size is still 1.6x times the median value but the NIR flare has dimmed.  

This shift in the synchrotron thermal spectrum at~$t$~=~37,460~GM/c$^3$ and~$t$~=~37,510~GM/c$^3$ causes~$\alpha_{submm}$ to become negative compared to the quiescent and peak NIR total flux times. We calculate~$\alpha_{submm}$~=~-0.84 at the peak sub-mm size and~$\alpha_{submm}$~=~-0.49 at~$t$~=~37,510~GM/c$^3$. 

\section{Behavior of Observables Across All Simulations}

\subsection{NIR Total Flux and Sub-mm Size}\label{sec:allsim_fNIR_s230}

\begin{figure}
        \includegraphics[width=\columnwidth]{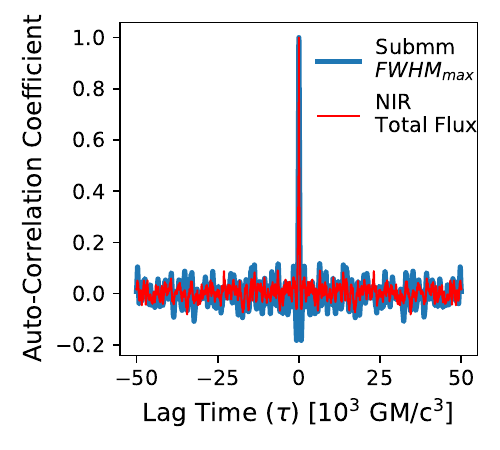}
        \caption{The auto-correlation functions for the NIR Total Flux (red) and the the sub-mm maximum size increase (light blue) over the time 10,000~GM/c$^3$ to 96,310~GM/c$^3$ for the fiducial high-spin, prograde simulation. Besides the expected peak at~$\tau$~=~0, there is no significant secondary peak for a~$\tau$ range of from 0~GM/c$^3$ to 50,000~GM/c$^3$, just over half the simulation time. Therefore, there are no significant periods of recurrence for both the NIR Total Flux and the sub-mm size increases.}
        \label{fig:auto_cc}
\end{figure}

\begin{figure}
        \centering
    \includegraphics[width = \columnwidth]{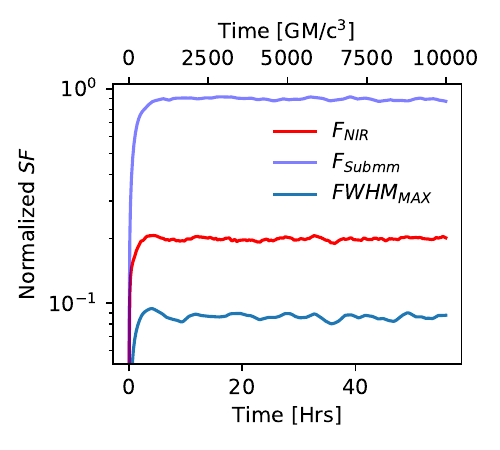}
    \caption{The structure function ($SF$, Equation~\ref{eq:SF}) for the total NIR flux~$F_{NIR}$, total sub-mm flux~$F_{submm}$, and the sub-mm max size increase~$FWHM_{MAX}$ plotted a log scale for the fiducial high-spin, prograde simulation. Each curve is normalized by its mean value. The structure function for~$F_{NIR}$ and~$F_{submm}$ reach a plateau value at~$\sim$5~hours. The structure function for the sub-mm ~$FWHM_{MAX}$ shows some variation over after reaching a peak value but still stays near the mean, normalized value of 0.18.}
    \label{fig:a9375_SF}
\end{figure}

By running the high-spin, prograde simulation to 96,310~GM/c$^3$, we were able to characterize the long-term behavior of the sub-mm and NIR total flux, centroid motion, and size. The auto-correlation functions for both the NIR total flux and the sub-mm $FHWM_{MAX}$ (Figure~\ref{fig:auto_cc}) show no prominent peaks besides~$\tau$~=~0, demonstrating a lack of periodicity. The structure functions~$SF$ for the NIR total flux, sub-mm total flux, and the sub-mm~$FWHM_{MAX}$ (Figure \ref{fig:a9375_SF}) all reach an asymptotic value $\tau_{SF}$ near 1,000~GM/c$^3$ (5~hours for Sgr A*). Since the NIR total flux curve $\tau_{SF}$ is less than the time between magnetic flux eruptions, the NIR flares are likely independent from one another. 

\begin{figure*}
    \centering
    \includegraphics[width = \textwidth]{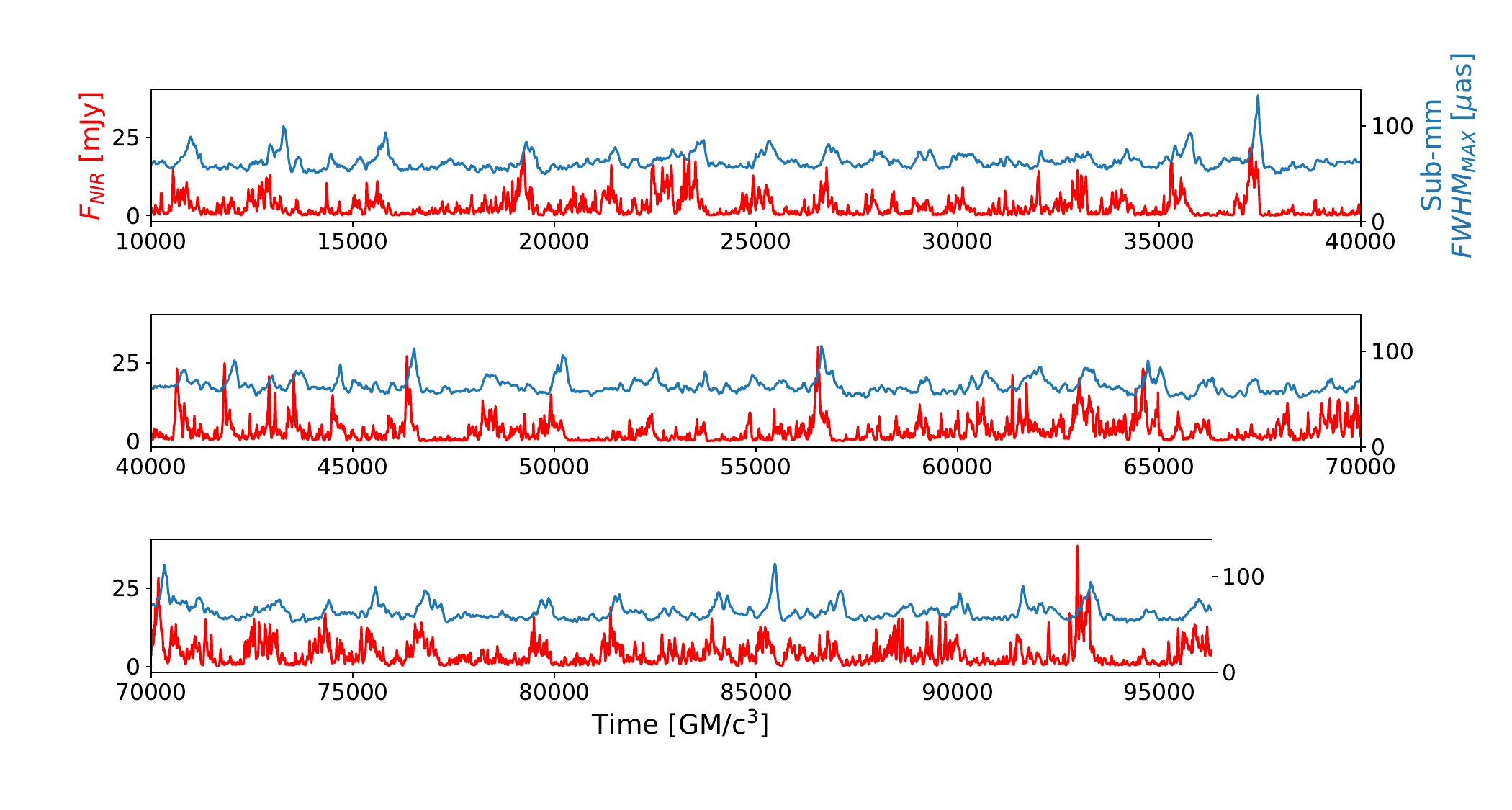}
    \caption{The Total NIR Flux and sub-mm~$FWHM_{MAX}$ for the full (10,000~GM/c$^3$ to 96,310~GM/c$^3$)~high spin, prograde simulation. Even at late times, there are many examples of significant size increases following a period of NIR flares. Some smaller amplitude flares are able to produce large size increases while large amplitude flares can produce less extreme size increases. As explored in the Discussion, the actual size increase observed is dependent on the geometry of the low density, high temperature bubble. The size increases generally have a single peak with similar rising and falling behavior that corresponds to the ejection and then re-accumulation of material near the horizon, respectively. However, the NIR flares demonstrate a variety of morphologies. Some flares exhibit a single, extreme peak followed by lesser peaks (e.g. times $\sim$57,000~GM/c$^3$ and $\sim$92,500~GM/c$^3$) while other flares show several medium amplitude peaks in a row (near 23,000~GM/c$^3$ and 89,000~GM/c$^3$).}

    \label{fig:fNIR_s230_alltime}
\end{figure*}

\begin{figure*}
    \includegraphics[width = .75\textwidth]{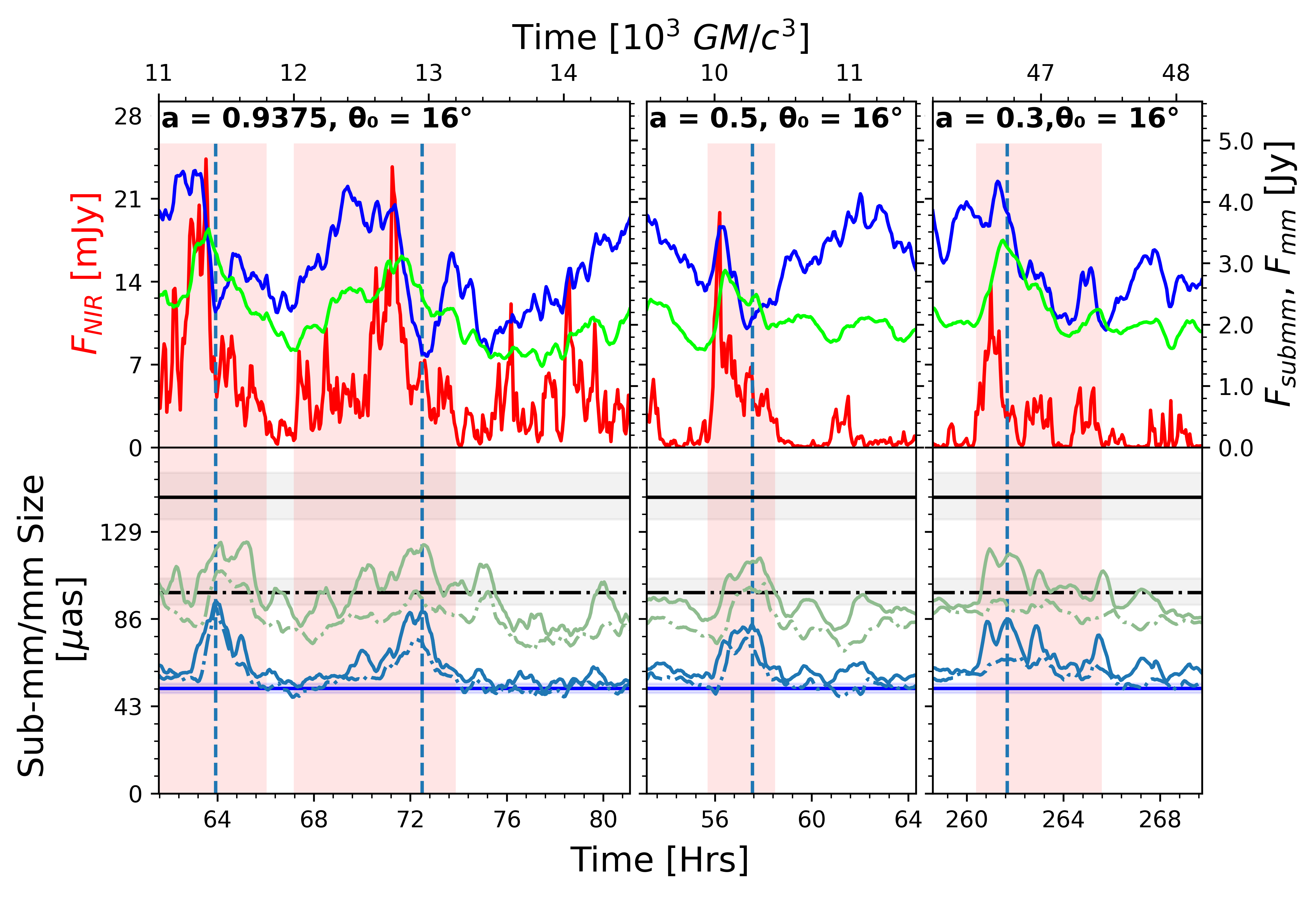}
    \includegraphics[width = \textwidth]{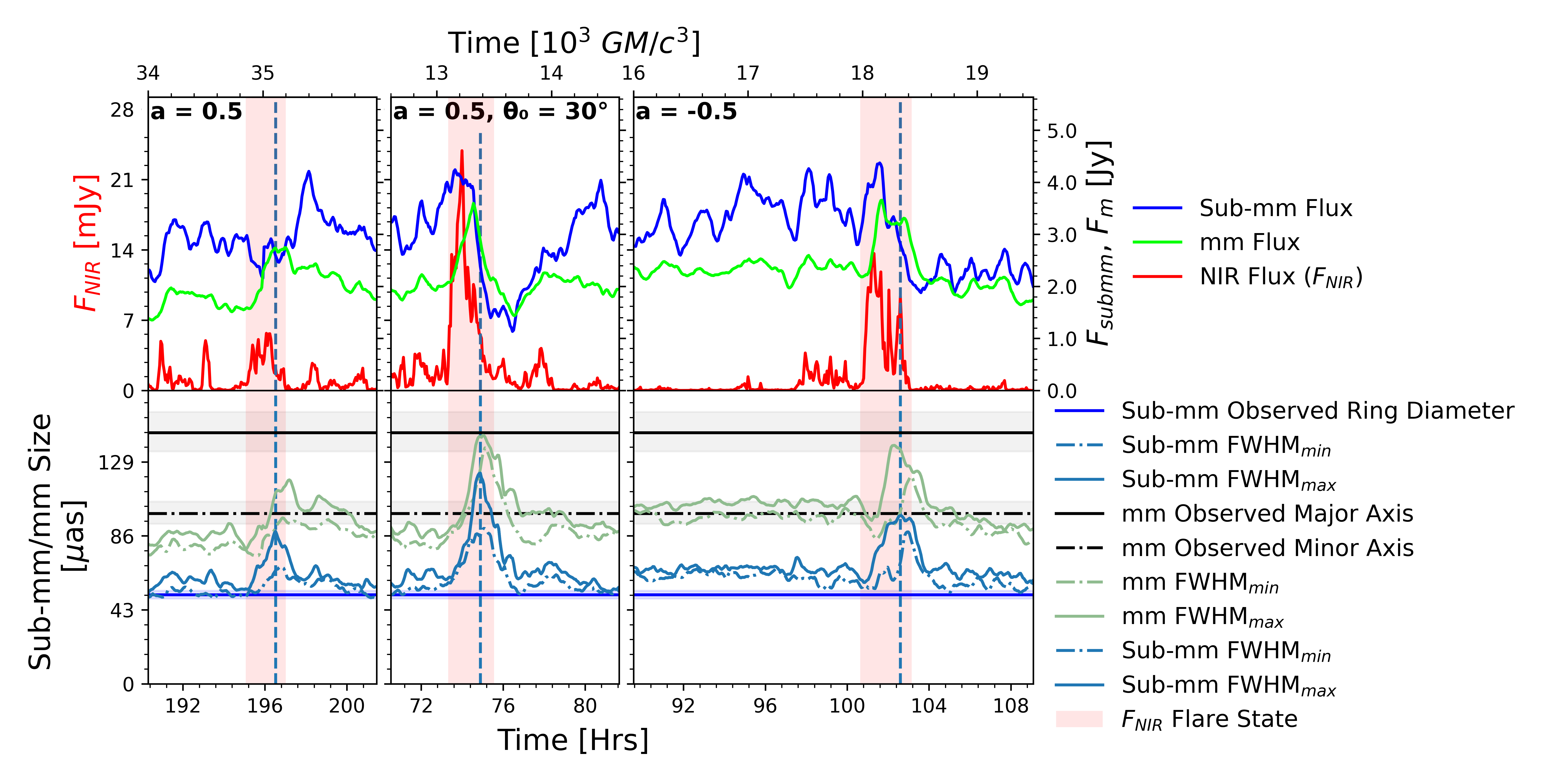}
    \caption{Examples of a NIR flare followed by a sub-mm and mm emission region size increase for every simulation studied in this work (with the exception of the fiducial simulation already shown in Figure \ref{fig:a9375_prograde_example}). Every column is a selected interval of time from a different simulation; the spin $a$ and initial tilt angle $\theta_0$ of the simulation are written in bold in each top left corner. The light red region highlights the interval of time where the NIR total flux is in a flare state. The light blue, dashed line aligns with the time where the sub-mm~$FWHM_{MAX}$ is at a maximum. The black, horizontal solid and dashed lines are the observed median intrinsic (de-scattered) mm (86~GHz) major and minor axis, respectively, from~\citet{Issaoun2021}. The grey, shaded region corresponds to uncertainty (95\% confidence interval) of the mm observed size. The blue, horizontal line represents the observed sub-mm size and the light blue shaded region is the boundary of the 68 \% confidence interval of the observed sub-mm size \citep{EHTsgrAstar_1}.  In all the examples shown, the sub-mm size increase peak occurs a few~minutes before the mm size increase peak. All simulations produce the observed sub-mm size during quiescent periods, but both the mm size and eccentricity is lower than the observed value.}
    \label{fig:allsims_posexamples_230size_86ghsize}
\end{figure*}

\begin{figure*}
    \centering

    \includegraphics[width = \textwidth]{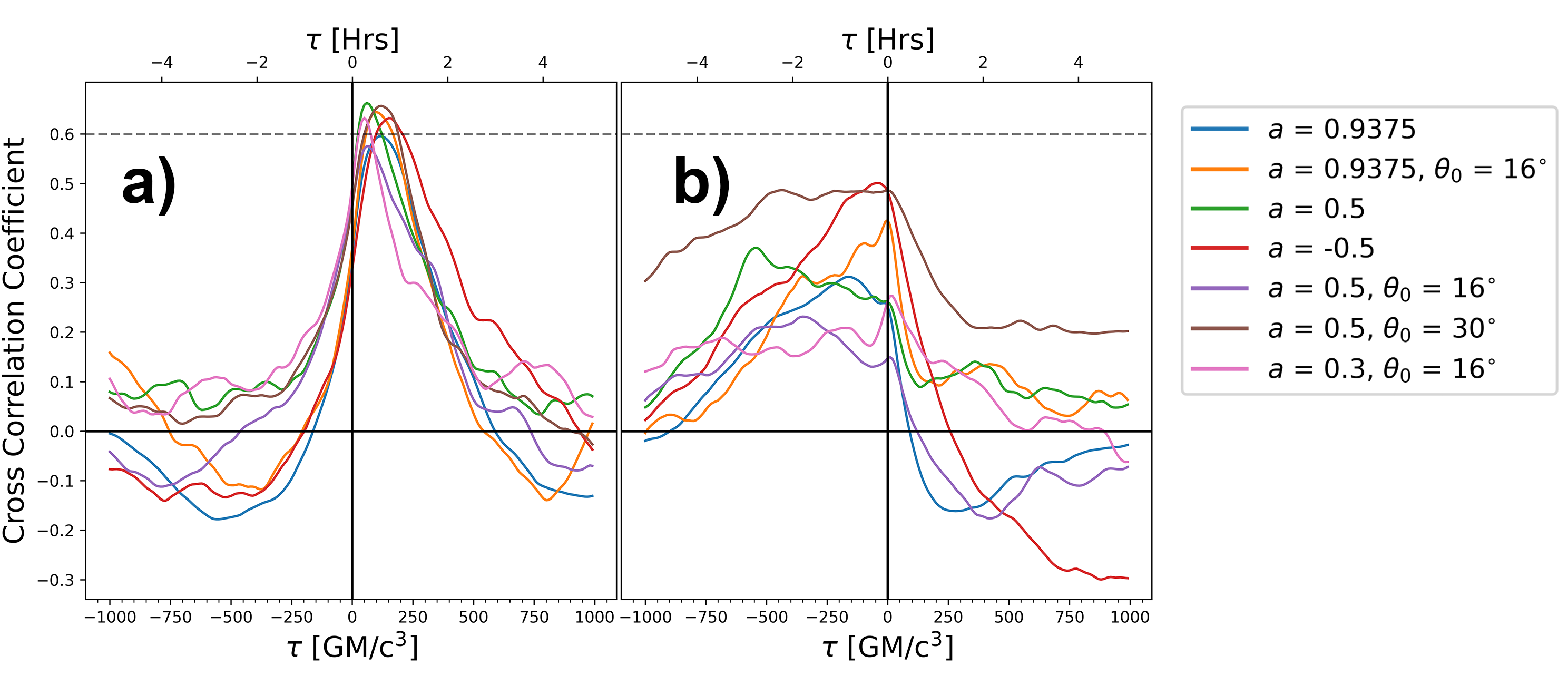}

        \caption{\textbf{a)} The cross correlation coefficient as a function of lag time~$\tau$ between the NIR total flux and the sub-mm~$FWHM_{MAX}$ for all simulations studied in this paper. The dotted grey line represents a cross-correlation coefficient value of 0.6. Regardless of initial tilt angle~$\theta_0$ or spin~$a$, all simulations demonstrate a strong, singular cross-correlation coefficient peak of about 0.6 and a characteristic time delay $\tau^*$ between 20 - 60 minutes. Simulations with low spin and low initial tilt have shorter peak lag times than those with high spins or moderate to high initial tilt angles. \textbf{b)} The cross correlation coefficients between the NIR and the sub-mm total flux show both wider peaks and smaller peak values than their counterparts between the NIR total Flux and sub-mm~$FWHM_{MAX}$. There is also greater variability of the cross correlation peak values, the peak lag times, and the peak widths across our parameter space for the NIR and sub-mm total flux versus NIR total flux and sub-mm~$FWHM_{MAX}$. Therefore, a correlation between the NIR and sub-mm total flux exists but is weaker and less consistent than the correlation between the NIR total flux and sub-mm~$FWHM_{MAX}$. }
        
        \label{fig:cc}
\end{figure*}

Significant increases in the sub-mm~$FWHM_{MAX}$ follow many NIR flares during the high-spin, prograde simulation run~(Figure~\ref{fig:fNIR_s230_alltime}). This relationship, however, is not unique to our fiducial simulation and persists across our entire parameter space. Figure~\ref{fig:allsims_posexamples_230size_86ghsize} shows a positive example of a sub-mm size increase following a NIR flare for every simulation studied. The cross-correlation coefficient between the NIR total flux and the sub-mm~$FWHM_{MAX}$ for every simulation has a single, significant peak with a value near 0.6, a positive characteristic time delay~$\tau^* <$~1~hour, and peak width of~$\sim$3~hours (Figure~\ref{fig:cc}). In our convention, a positive $\tau^*$ means that the NIR total flux curve's behavior precedes the sub-mm~$FWHM_{MAX}$ curve's behaviour.

However, we do find a dependence between the black hole spin and accretion flow orientation on~$\tau^*$: low-spin and low initial tilt angle ($\theta_0$) simulations produce the shortest $\tau^* \sim$~20~minutes while increasing either the spin or $\theta_0$ increases the value of $\tau^*$ to 40~minutes~-~50~minutes. 

In all simulations studied, the NIR total flux, unlike the the sub-mm $FWHM_{MAX}$, is intrinsically noisy. This noise reduces the value of the peak cross-correlation coefficient between the NIR total flux and the sub-mm $FWHM_{MAX}$. Performing a moving average of only the NIR total flux for the high-spin, prograde simulation with a window size between 120~GM/c$^3$~-~240~GM/c$^3$ (40~minutes~-~80~minutes) preserves the periods of significant flaring while reducing the NIR total flux noise. When we correlate this averaged NIR total flux curve with the sub-mm $FWHM_{MAX}$, the peak cross-correlation coefficient between the two curves increases from 0.6 to 0.7 - 0.8.

\subsection{The Probability of a Sub-mm Size Increase}

Although the cross-correlation coefficient illustrates the relationship between the NIR total flux and the sub-mm size increases, the curve alone does not assign a probability of a size increase given a NIR total flux flare. To derive those probabilities, we first define an \textit{observation window} to be a time-ordered set of all observables within a fixed time interval. Each observation window is indexed by its start time. We define the random variable $X$ as a Boolean-valued outcome of the following trial: a success is if there is at least one point within an observation window of the sub-mm $FWHM_{MAX}$ above some threshold value $\alpha_{FWHM}$, where $\alpha_{FWHM}$ is normalized by the median sub-mm $FWHM_{MAX}$, and a failure otherwise. We then define $f_{NIR}$ as the total NIR flux normalized by its median value (~$f_{NIR} = F_{NIR}/m(F_{NIR})$). The Boolean random variable $Y$ is then defined to be true if the total flux during an observation window ever exceeds the threshold value $f_{NIR}$ and false otherwise. 

We treat both $X$ and $Y$ random variables as independent and identically distributed. We then calculate the conditional probability $P(\alpha_{FWHM} > X | f_{NIR} > Y)$ for two fixed duration observation windows: 50~minutes  and 100 minutes. 

\begin{figure}
    \includegraphics[width=\columnwidth]{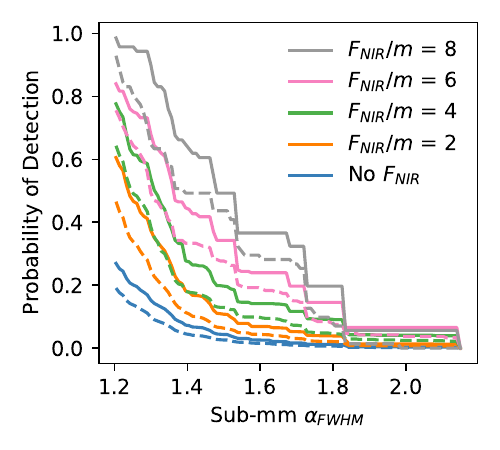}
    \caption{The probability of detecting at least one point in the fiducial high-spin, prograde simulation where the normalized sub-mm~$FWHM_{MAX}$ is above some threshold $\alpha_{submm}$ given that the NIR total flux is also elevated above the normalized flux $f_{NIR}$ at some point during the same time interval (observation window). We define the sub-mm $\alpha_{FWHM} = FWHM_{MAX}/m(FWHM_{MAX}) = FWHM_{MAX}/61.5 \mu \textrm{as}$, where $m(FWHM_{MAX})$ is the median value of the sub-mm $FWHM_{MAX}$ curve. The quantity $f_{NIR} = F_{NIR}/m(F_{NIR})$ and the median value of the NIR total flux is $m(F_{NIR}) = 2.2$~mJy. The solid lines correspond to a long, 100~minute observation window and the dashed lines correspond to the short, 50~minute observation window. The blue lines correspond to~$P(\alpha_{FWHM} > X)$ and the other lines correspond to~$P(\alpha > X | F_{NIR}/m > N)$ for an integer~$N$. Even restricting observations to periods of minor (x2) elevated NIR total flux significantly increases the chances of a positive sub-mm size increase detection. The stronger the flare, the higher the chance of detection. It is possible for the probability of a sub-mm size increase to reach 95\% if it occurred within an observation window containing a NIR flare that increased by a factor of x8 from its median value (such flares have already been observed, see~\citet{GRAVITY2020_NIRFLUXdist,Do_ginatNIRfluxflare, Witzel2018_NIRsurvey}).}
    \label{fig:conditional_P}
\end{figure}

Figure \ref{fig:conditional_P} shows these conditional probabilities for the fiducial high-spin, prograde simulation. An increase in NIR flux during an observation window greatly improves chances of detecting a sub-mm size increase during the same window. Even restricting observations to periods of minor (x2 the median value) elevated NIR total flux significantly increases the chances of a positive sub-mm size increase detection. For NIR flares that reach 8x the median NIR total flux, the chance of detecting at least a moderate size increase $>$ 90\%. Without considering the NIR total flux, the chance of detecting a moderate size increase is $\sim 20$\%. 

\subsection{The Sub-mm and NIR Flux}\label{sec:allsim_flux}

Unlike the consistent behavior of the cross-correlation coefficients between the sub-mm size and the NIR total flux, the cross-correlation coefficient between the sub-mm \textit{total flux} and NIR total flux varies significantly across the simulation parameter space~(See Figure~\ref{fig:cc}(right)). The resulting curves show a variety of peak values and values of the characteristic time $\tau^*$. The peak of the cross correlation coefficients are all near or significantly less than 0.5 and are also wider than their counterparts between the sub-mm size and the NIR total flux. Although we do find correlations between the sub-mm total flux and NIR total flux, the correlations between the sub-mm \textit{size} and the NIR total flux is both stronger and more consistent than the former across our parameter space.

\subsection{The NIR Centroids}\label{sec:allsim_centroids}

The NIR centroid motion varies considerably between flaring states and quiescent states. During quiescent periods, the NIR centroid stays near a median value with several ``jumps'' around the black hole. However, during flaring periods, the NIR intensity map is dominated by a bright region which moves in a smooth, continuous motion. For some extreme NIR flares, this continuous motion forms a quasi-circular pattern on the sky.

This qualitative behavior occurs in all simulations studied. However, the effective diameter of the centroid path during different NIR flares varies considerably both during the course of a single simulation and across the simulation parameter space. The example from the fiducial simulation explored in Section~\ref{sec:ex_a9375_cNIR_s86_s230} shows the largest (46~$\mu \textrm{as}$) effective diameter an NIR centroid makes on the sky out of all simulations studied here. Other quasi-circular centroid motions during NIR flares from the same simulation show smaller effective diameters of~$\sim$25~$\mu \textrm{as}$ and periods of 80-120~minutes. With the exception of the~$a$~=~0.3,~$\theta_0$~=~16$^{\circ}$ simulation, which only produces centroid motion with a diameter of 17~$\mu\textrm{as}$, the largest centroid motion diameters from other simulations explored range from~$\sim 20 \mu \textrm{as} - 25 \mu \textrm{as}$. 

In the example from our fiducial simulation, the sub-mm centroid also traces a quasi-circular orbit on the sky while the NIR centroid makes a similar orbit. However, we find many examples of NIR flares across our parameter space that exhibit the quasi-circular NIR centroid motion without similar sub-mm centroid motion. However, these same NIR flare examples  
do display significant sub-mm size increases. 
 
One interesting 9~hour period of time during the prograde~$a = -0.5$ simulation contains 10~consecutive NIR flares. During this period, the NIR centroid makes several loops around the sky rather than the usual one or two loops.

We can also use the cross-correlation coefficient to show that the quasi-circular motion NIR centroid seen during NIR flares does not continue during quiescent periods. We treat the $x$ and $y$ centroid components, $\bar{I}_x$ and $\bar{I}_y$, as separate curves and find the cross-correlation coefficient between them,~$\mathcal{CC}[\bar{I}_x, \bar{I}_y]$ for short $\sim$2,000~GM/c$^3$ (11~hour) time intervals that contain a significant NIR flare and for the entire duration of the simulation. If the centroid traced a perfect ellipse on the sky with frequency~$\omega$,then the maximum of the cross correlation coefficient~$\mathcal{CC}[\bar{I}_x, \bar{I}_y]_{MAX} = 1$ at~$\tau =\sqrt{2}/2\omega$. For the short intervals of time, we find examples that reach~$\mathcal{CC}[\bar{I}_x, \bar{I}_y]_{MAX}$ of 0.6 - 0.8. However, the values of~$\mathcal{CC}[\bar{I}_x, \bar{I}_y]_{MAX}$ for the full duration of each simulation in our parameter space is much lower at 0.2 - 0.4. The upper-end of this interval comes from the $a = -0.5$ simulation where there is a 9~hour period of NIR flaring that results in many orbits of the NIR centroid around the black hole. 

\section{Discussion}\label{Discussion}

In this work, we show that magnetic flux eruptions cause a low density, high temperature magnetized bubble to form in the inner accretion flow. This bubble orbits around the black hole and breaks into smaller regions before reaching large ($> 30 r_g$) radii. The low density of the bubble as well as temperature increases in the $\sigma < 1$ regions of the accretion flow lead to significant (1.5x - 2x) increases in the sub-mm and mm emission region. These size increases follow NIR flares by $\sim$ 1~hour for the high-spin, prograde simulation. The mm size increase trails behind the sub-mm size increase by a few minutes, likely due to the fact that the mm emission region extends to larger radii compared to the sub-mm emission region. The sub-mm size increase coincides with a sudden drop in the sub-mm total flux and can cause the sub-mm spectral index to become more negative. This relationship between NIR flaring and sub-mm/mm size increases occurs over a parameter space of black hole spin and initial tilt angle: parameters currently unconstrained for Sgr A*. 

We showed that magnetic flux eruptions produce quasi-circular centroid tracks for simulations over a variety of black hole spin values; the largest diameter track occurred during the high spin case. We identified several examples of NIR flares where the NIR centroid follows a quasi-circular path but the sub-mm centroid does not. The sub-mm and NIR centroid motions are weighed by nearly complimentary regions in their respective emission maps~(See Figure~\ref{fig:images}). There are also many sources that can effect the position of the sub-mm centroid: the location of the shadow caused by the low density ``bubble,'' additional heating outside the ``bubble,'' and the anisotropic geometry of the accretion flow itself. Therefore, NIR centroid motion does not guarantee similar sub-mm centroid motion.

We also considered a single, high spin~($a = 0.9375$) SANE simulation with~$\theta_0 = 16^{\circ}$. After running the simulation for 25,000~GM/c$^3$, we found one example of a small NIR flare ($<$ 1~mJy) and no examples of significant sub-mm size increases. Therefore, the extreme variability in sub-mm size may be a unique feature of magnetic flux eruption events that occur in MAD simulations.

In our work, the accretion rate is fixed over 5,000~GM/c$^3$ periods of time by matching the calculated, average sub-mm total flux to its observed value ($\sim$3 Jy) for Sgr A* . In the real Sgr A* system, stellar winds from the~$\sim$30 Wolf-Rayet stars that populate the galactic center provide a significant source of accretion material~\citep{Ressler2018, Ressler2019, Ressler2020}. 

\subsection{Electron Heating, Non-Thermal Electrons, and $\sigma$-cut}\label{eheating_sigma_nonthermal}

Self-consistently evolving the electron temperatures requires a choice of electron heating prescription. 

In their study of sub-mm emission from a magnetic flux eruption, \citet{jia2023} applied the $R_{high} - R_{low}$ electron heating
model \citep{moscibrodzka2016}, where $R \equiv T_p/T_e$. They demonstrated that, for high $T_e$ models, setting $R = 1,10$ and ray-tracing images with the $\sigma > 1$ regions excluded produced a steady decline in the sub-mm total flux and did not produce significant sub-mm size increases. However, setting $R = 100$ caused an increase in the sub-mm size. The sub-mm total flux also increased in the latter case before fading.

For this study, we chose the electron heating prescription W18 ~\citep{Werner2018} because it was the most successful model in matching mm to NIR observations of Sgr A* \citep{Dexter2020_survey} and producing NIR centroid motions. Models of particle-in-cell turbulent electron heating and heating during reconnection for an electron/ion plasma is still an active field of study. Future results in this field will provide new prescriptions and more accurate predictions of observables from our models. Here, we also apply the W18 electron heating fraction uniformly regardless of whether reconnection is occurring. Future studies could account for spatial variation in the heating mechanism, instead of solely the fluid magnetization.



In our model, the electrons in the accretion flow are thermal. Relativistic reconnection  \citep{Sironi2014, French2023, hayk_2023_secondaryplasma} the magnetic Rayleigh-Taylor instability \citep{Zhdankin2023_RT} are possible sources for non-thermal particles in the accretion flow. Non-thermal particles can significantly increase the NIR total flux while having a less pronounced effect on the sub-mm/mm observables~\citep{DoddsEden2009, Scepi2022}. Non-thermal particles may also be the cause of the large mm size on the sky compared to what is calculated in this work. Non-thermal emission may also decrease the true NIR flux if electrons actually emit at higher frequencies (such as the X-ray).

In the images presented in our Results, we also excluded the high $\sigma > 1$ regions of the accretion flow ($\sigma\textrm{-cut} = 1$). As shown in Figure \ref{fig:HARM_var}, large regions of the innermost accretion flow ($r \lesssim 10 r_g$) reach $\sigma > 1$ values during a magnetic flux eruption. Setting $\sigma\textrm{-cut} = 10$ includes most of the highly magnetized bubble in the image while still excluding the jet.

Increasing $\sigma\textrm{-cut}$ to 10 only produces a slight (2 $\mu\textrm{as}$) increase in the sub-mm size during quiescence compared to its $\sigma\textrm{-cut} = 1$ counterpart. The $\sigma\textrm{-cut} = 10$ images from our fiducial example still show a sub-mm size increase reaching 2x its median value. The gap that forms in sub-mm intensity maps also persist in the high $\sigma$ image, demonstrating this gap is due to the drop in density in the region rather than an artificial omission of emission stemming from our choice of a low $\sigma$ cut. The difference in major axis of the peak sub-mm size between $\sigma\textrm{-cut} = 10$ and $\sigma\textrm{-cut} = 1$ is negligible ($<$ 0.002 $\mu\textrm{as}$) while the minor axis only increases in the higher $\sigma\textrm{-cut}$ case by $\sim$ 3 $\mu\textrm{as}$. The $\sigma\textrm{-cut} = 10$ NIR images produce roughly double the total NIR flux compared their $\sigma\textrm{-cut} = 1$ counterparts (see \citet{Dexter2020_NIRcentroid} for further discussion on the choice of $\sigma\textrm{-cut}$ and its effect on the NIR image.)

\subsection{Electron Cooling} 

 Electrons radiating at a frequency~$\nu$ lose energy and cool over a timescale~$t_{syn} \propto \nu^{-1/2}$. We do not include synchrotron cooling in our simulations. Recent results suggest that radiative cooling can effect the total luminosity and flux peaks in both the millimeter and NIR wavelengths~\citep{Yoon2020_radGRMHD}. Non-thermal electrons in the accretion flow will also radiate at high energies and cool. Particle-in-cell studies of electron/positron pair reconnection show that radiative models produce different power laws than their non-radiative counterparts \citep{Werner2019_ICcooling, sironi2020_radPIC}.

 For the NIR photons,~$t_{syn, NIR}$~$\sim$10~minutes. Because ~$t_{syn, NIR}$ is less than both the flare timescale and the duration of the quasi-circular NIR centroid orbits, cooling may have a significant impact on our results. 

The  sub-mm synchrotron cooling time, by contrast, is ~$\sim$4~hours and much longer than the time scale for the sub-mm size increases ($\sim 2$~hours) and sub-mm centroid motion. Also, the sub-mm total flux value is manually fixed. However, the sub-mm size increases are highly dependent on the hot, low density bubble from forming in the first place. If the hot NIR electrons cool too quickly in the real accretion flow, the bubble may dissipate more rapidly and not lead to such large sub-mm size increases. 

\subsection{MAD GRMHD Models and Simulation Resolution}

Our models are subject to the same numerical computation constraints found in other magnetically dominated GRMHD simulations. Strongly magnetic regions are difficult to evolve but are present in the low density ``bubble'' that drives size increases and centroid motion.

Extreme resolution (5376×2304×2304) GRMHD simulations of black hole accretion flows produce magnetic flux eruption events that show reconnection features such as plasmoids and X-points~\citep{giantGRMHD}. However, the extreme resolution simulations produced lower reconnection rates than its lower resolution counterpart because of the  larger numerical resistivity in the latter case \citep{giantGRMHD}. Still, magnetic reconnection in GRMHD simulations is generally slower than in particle-in-cell (kinetic) simulations because of pressure anisotropies that are not well captured in MHD \citep{Bessho2005_2Dpressureinisotrpy,Bransgrove2021}. Therefore, the duration of the magnetic flux eruption events in this work may be overestimated or underestimated.

Numerical diffusivity likely effects the boundary of the highly magnetized bubble as it orbits around and away from the black hole. The temperature increases directly outside the $\sigma > 1$ region leads to changes in the sub-mm size and NIR total flux. Studying changes in sub-mm and NIR images from higher resolution GRMHD simulations may demonstrate if the sub-mm size increases (and their relationship to the NIR total flux shown in this work) depend strongly on our choice of resolution. At lower resolution (1/2 to 3/4 the number of cells compared to the resolution explored here),~\citet{Dexter2020_NIRcentroid} found that NIR~hot spot formation and rotation around the black hole still occurs.

Despite the success of MAD accretion flows in explaining a variety of Sagittarius A* observations, there still exists some tension between simulated observables and real observations. \citet{EHTsgrastar5} concluded that, although MAD models with inclination angles~$< 30^{\circ}$ were preferred, no model in their library of static images met all their observational constraints. 

\subsection{Dynamic Imaging with the EHT}

\begin{figure*}
    \centering
    \includegraphics[width= \textwidth]{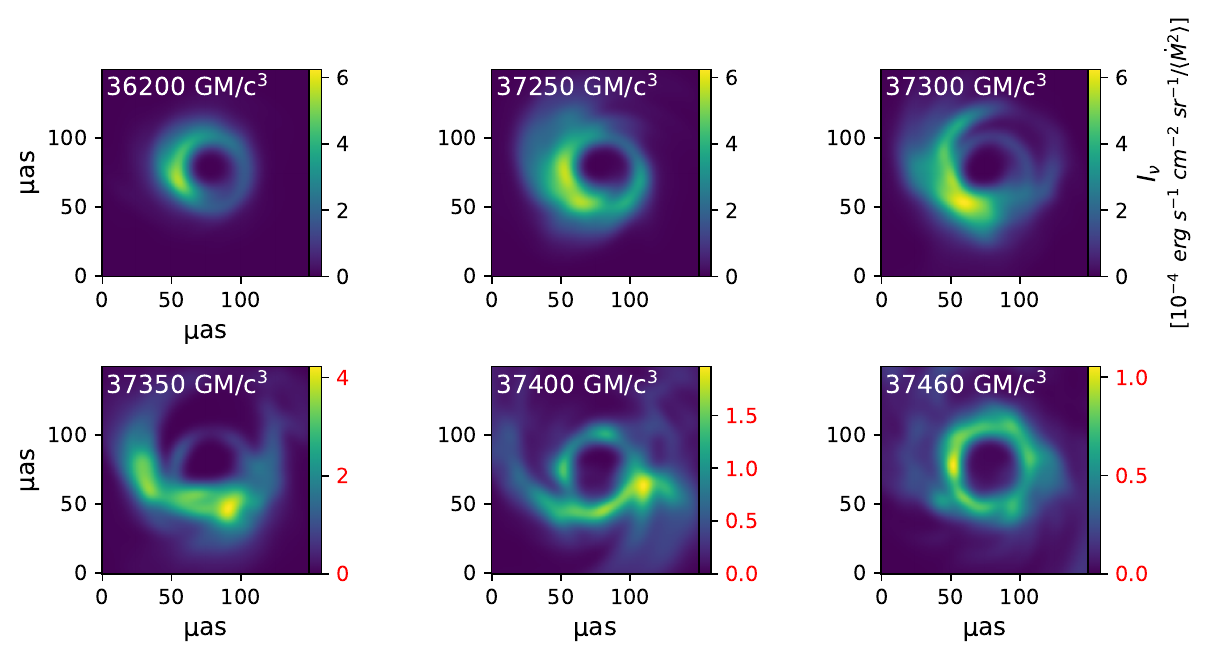}
    \caption{Simulated sub-mm (230~GHz) ground truth images using a 20~$\mu \textrm{as}$ FWHM Gaussian kernel for the fiducial NIR flare example shown in Figure~\ref{fig:a9375_prograde_example} and~\ref{fig:a9375_prograde_example_centroids}. The field of view is 150~$\mu \textrm{as}$; note that the color bar scale differs for the bottom row of images. The image intensity is plotted on a linear scale. The bright spot in the emission region clearly rotates around the black hole and the emission region also undergoes prominent morphological changes.}
    \label{fig:EHTeyes_examplemosaic}
\end{figure*}

EHT imaging algorithms are sensitive to ring morphologies~\citep{EHT_sgrAstar_3}. In order to recreate the resolution of the constructed EHT image from interferometric data, we blur the 150~$\mu \textrm{as}$ x 150~$\mu \textrm{as}$ region centered on the black hole in our sub-mm simulated images with a 20~$\mu \textrm{as}$ FWHM Gaussian kernel. We focus on the blurred sub-mm before and during the fiducial NIR flare~(Figure \ref{fig:EHTeyes_examplemosaic}). Not only do we find that the resulting structure is ring-like in all images, but we can clearly see the sub-mm hotspot rotating around the black hole center. The effective shadow in the middle of the ring is also larger during a sub-mm size increase than during quiescence. Further analysis with dynamic imaging algorithms is required to confirm whether this variability can be reconstructed with current EHT or ngEHT~$u,v$ coverage.

\subsection{Inclination Angle and The Sub-mm/mm Peak Size}

\begin{figure}
    \centering
    \includegraphics[width=\columnwidth]{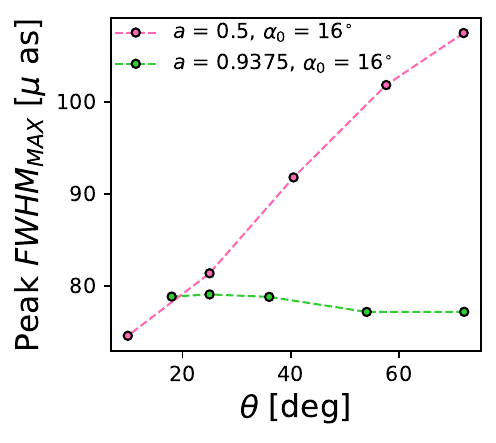}
    \caption{Two extremes of inclination angle dependence on the sub-mm (230~GHz) maximum size. Both curves show a size increase that occurred after a flaring event: the green line is taken from the~$a = 0.9375$,~$\theta_0 = 16^{\circ}$ simulation and the pink line comes from the~$a = 0.5$,~$\theta_0 = 16^{\circ}$ simulation. In the pink example, the peak value of the sub-mm maximum size curve grows linearly with angle. By contrast, in the green example, the size increase stays relatively flat for a variety of inclinations. In both examples, the peak size increase is significantly elevated from its median value for all inclination angles.}
    \label{fig:incl_angle_dep}
\end{figure}

Observations of Sgr A* are broadly consistent with a low inclination angle~\citep{EHTsgrastar5, GRAVITY2018} and motivate our choice of~$i$~=~25 \degree. However, the exact value of the inclination angle to Sgr A* is still unclear. We varied the inclination angle for the time interval containing NIR flares from several simulations and studied the effect on the peak sub-mm size increase. The peak sub-mm size increase can remain prominent for very low (10$^{\circ}$) and very high (73$^{\circ}$) inclination angles. However, the exact relationship between the sub-mm peak value and the inclination angle can vary dramatically for different NIR flare examples. Figure~\ref{fig:incl_angle_dep} demonstrates two extremes of this relationship. In one example, the sub-mm peak size is significantly elevated from the median size value but remains flat as a function of inclination angle. In the second example, the sub-mm peak size grows roughly proportionally to the inclination angle. 
 
 Since 1.3~mm~(230~GHz) is near the synchrotron spectrum peak, the intensity map at 1.3~mm is sensitive to the underlying density distribution of the gas. The actual shape of the the low-density, high temperature region resulting from the magnetic flux eruption event varies considerably. Therefore, the shape of the 1.3~mm emission region will also vary. Although the examples come from two simulations with different spin $a$, it is unclear if $a$ has an strong effect on this relationship between inclination angle and peak sub-mm~size. 
 
\subsection{Comparison to the Observed 86~GHz/230~GHz Quiescent Size and Observed NIR Centroid}\label{discuss:sizes}

All simulations explored in this work reproduce the observed sub-mm EHT size during quiescent periods, although the median sub-mm size can be slightly ($\sim$15\%) larger than the observed sub-mm EHT size. 

By contrast, there is tension between our modeled mm size and the observed mm size. Our simulations consistently produce a maximum quiescent mm size that is about 30\% lower than the observed value from the most recent analysis of ALMA data from April 3rd, 2017~\citep{Issaoun2021}. The quiescent eccentricity of the mm (86~GHz) emission is near zero in these simulations whereas the observed value is~$\sim$0.7. Interestingly, the predicted size and eccentricity matches the ALMA data well during a mm size increase following a NIR flare~(See Figure~\ref{fig:a9375_prograde_example} from Section~\ref{sec:ex_a9375_cNIR_s86_s230}). Future observations may shed light on the true distribution of the mm~size and its variability following NIR flares.

The fiducial NIR flare studied here still produces a NIR centroid track diameter that is a factor of 3 smaller than observed diameter from the GRAVITY 2018 measurement~\citep{GRAVITY2018}. With an effective diameter of~$46 \mu\textrm{as}$, the NIR centroid motion from the featured example is also the largest in our parameter space. NIR centroid motions with diameters of~$\sim 15 \mu \textrm{as}~-~25 \mu \textrm{as}$ are more common in our simulations but 7x - 8x smaller than the observed value. Sources of non-thermal NIR emission missing from our model, such as particle acceleration and synchotron cooling, can account for the discrepancy in the NIR orbit size \citep{Ball_plasmoidmodel_NIR}. 

\section{Conclusions}

GRMHD simulations of magnetically arrested disks undergo magnetic flux eruptions that power NIR flares and may explain the NIR centroid motion observed by GRAVITY. Additionally, the same class of simulations predict significant and frequent sub-mm and mm size increases that occur within a 2 hour window after a NIR total flux peak. The models presented here make a clear, qualitative prediction within reach of the Next Generation Event Horizon Telescope.

\section*{Acknowledgements}

The authors thank D. Uzdensky, A. Hankla, P. Dhang, and T. McMaken for their helpful comments. This work was supported in part by NASA Astrophysics Theory Program grant 80NSSC20K0527, Fermi award 80NSSC21K2027, Chandra award TM3-24003X, and by an Alfred P. Sloan Research Fellowship.

\section*{Data Availability}

Simulated images and averaged simulation data products used here will be shared on reasonable request to the corresponding author.



\bibliographystyle{mnras}
\bibliography{example} 






\bsp	
\label{lastpage}
\end{document}